\documentclass[12pt]{article}
\pdfoutput=1
\usepackage{bbm}
\usepackage{color}
\usepackage{diagbox}
\usepackage{authblk}
\usepackage{latexsym}
\usepackage{epsfig,amssymb,euscript, mathrsfs,cite}
\usepackage{amsmath}
\usepackage{mathrsfs} 
\usepackage{enumerate}
\usepackage{slashed}
\definecolor{MyDarkBlue}{rgb}{0.15,0.15,0.45}
\usepackage[linktocpage=true]{hyperref}
\hypersetup{
colorlinks=true,
citecolor=MyDarkBlue,
linkcolor=MyDarkBlue,
urlcolor=MyDarkBlue,
pdfauthor={},
pdftitle={},
pdfsubject={hep-th}
}

\newcounter{savefootnote}
\newcounter{symfootnote}
\newcommand{\symfootnote}[1]{%
   \setcounter{savefootnote}{\value{footnote}}%
   \setcounter{footnote}{\value{symfootnote}}%
   \ifnum\value{footnote}>8\setcounter{footnote}{0}\fi%
   \let\oldthefootnote=\thefootnote%
   \renewcommand{\thefootnote}{\fnsymbol{footnote}}%
   \footnote{#1}%
   \let\thefootnote=\oldthefootnote%
   \setcounter{symfootnote}{\value{footnote}}%
   \setcounter{footnote}{\value{savefootnote}}%
}

\usepackage{mathtools}

\newsavebox{\ns}
\newsavebox{\dbrane}
\newsavebox{\dbshort}

\def\be{\begin{equation}}
\def\ee{\end{equation}}
\def\bea{\begin{eqnarray}}
\def\eea{\end{eqnarray}}

\newcommand{\nn}{\notag \\}

\def\eq#1 { \begin{equation} #1 \end{equation} }

\newcommand\diff{\mathrm{d}}

\newcommand{\tr}{\mathrm{tr}}

\newlength{\sswidth}

\numberwithin{equation}{section}       

\begin{document}

\begin{titlepage}

\vfill

\begin{flushright}
Imperial/TP/2022/JG/02\\
\end{flushright}

\vfill

\begin{center}
   \baselineskip=16pt
   {\Large\bf Leigh-Strassler compactified on a spindle\symfootnote{No physicists were harmed in carrying out the research reported here.}}
  \vskip 1cm
Igal Arav$^1$,  Jerome P. Gauntlett$^2$\\
Matthew M. Roberts$^2$ and Christopher Rosen$^3$\\
     \vskip 1cm     
                          \begin{small}
                                \textit{$^1$Institute for Theoretical Physics, University of Amsterdam,\\
                                Science Park 904, PO Box 94485,\\
                                1090 GL Amsterdam, The Netherlands}
        \end{small}\\
        \begin{small}\vskip .3cm
      \textit{$^2$Blackett Laboratory, 
  Imperial College\\ London, SW7 2AZ, U.K.}
        \end{small}\\
             \begin{small}\vskip .3cm
      \textit{$^3$Crete Center for Theoretical Physics, Department of Physics, University of Crete,\\
71003 Heraklion, Greece}
        \end{small}\\
                       \end{center}
\vfill

\begin{center}
\textbf{Abstract}
\end{center}
\begin{quote}
We construct a new class of supersymmetric $AdS_3\times Y_7$ solutions of type IIB supergravity,
where $Y_7$ is an $S^5$ fibration over a spindle, which are dual to $d=2$, $\mathcal{N}=(0,2)$ SCFTs. 
The solutions are constructed in a sub-truncation of $D=5$, $SO(6)$ maximal gauged supergravity and they all lie within the anti-twist class. We show that the central charge computed from the gravity solutions agrees with an anomaly polynomial calculation associated with compactifying the $\mathcal{N}=1$, $d=4$ Leigh-Strassler SCFT on a spindle.

\end{quote}

\vfill

\end{titlepage}

\tableofcontents

\newpage

\section{Introduction}\label{sec:intro}

A fruitful way to engineer supersymmetric conformal field theories (SCFTs) is to compactify higher-dimensional 
SCFTs. An important paradigm \cite{Maldacena:2000mw} is to consider a SCFT on the product of flat spacetime 
with a compact manifold and then, in order to preserve supersymmetry,
switch on background magnetic fluxes which encapsulate a partial topological twist of the SCFT. In favourable situations, 
the system will then flow to a new SCFT at low energies. If the parent SCFT field theory has a large $N$ holographic dual 
one can study the resulting renormalisation group (RG) flow holographically by appropriately constructing dual supergravity solutions. In fact such holographic solutions
provide an important tool in establishing whether or not the compactified SCFT flows to a new SCFT in the IR.

Recently it has been appreciated that this well studied paradigm can be modified in two interrelated ways. Firstly, one can relax
the condition that the compact space is a manifold and instead consider orbifolds. In particular, starting with \cite{Ferrero:2020laf}, there has been 
considerable work studying SCFTs compactified on a spindle, a two-dimensional orbifold which is topologically a two sphere but with conical deficit angles at the north and south poles. 
Secondly, supersymmetry is no longer realised by the standard topological twist. For spindles with an azimuthal symmetry, which
is the class that has been studied, there are just two ways to preserve supersymmetry called the ``twist" and
the ``anti-twist" \cite{Ferrero:2021etw}, which are characterised by the $R$-symmetry flux through the spindle. The twist is in the same topological class as the standard topological twist but there are some differences: for example the spinors on the spindle that are associated with the preserved supersymmetry are no longer constant and chiral. The anti-twist on a spindle is a new way of preserving supersymmetry.

The analysis of \cite{Ferrero:2020laf} was in the context of $\mathcal{N}=1$, $d=4$ SCFTs which are dual to $AdS_5\times SE_5$ solutions of type IIB supergravity, that are then reduced on a spindle. It was shown that these give rise to $\mathcal{N}=(0,2)$, $d=2$ SCFTs 
that are dual to $AdS_3\times Y_7$ solutions of type IIB supergravity, first found in \cite{Gauntlett:2006af}, where $Y_7$ is a smooth seven-dimensional manifold consisting of a fibration of the five-dimensional Sasaki-Einstein manifold, $SE_5$, over the spindle $\Sigma$. 
These supergravity solutions, which are all in the anti-twist class, were constructed as $AdS_3\times \Sigma$ solutions of $D=5$ minimal gauged supergravity and then uplifted on $SE_5$ to type IIB. It is particularly interesting that for $SE_5$ in the regular class,
the orbifold singularities of the spindle are eliminated after uplifting to the type IIB solutions. 
For the specific case of $\mathcal{N}=4$, $d=4$ SYM, one can include
additional background magnetic fluxes on the spindle and the corresponding $AdS_3\times \Sigma$ solutions can be 
constructed using the $D=5$ STU theory  \cite{Hosseini:2021fge,Boido:2021szx,Ferrero:2021etw}.
The STU theory arises as a consistent truncation of type IIB on $S^5$ and has a bosonic content consisting of a metric, $U(1)^3$ gauge fields and two neutral scalars \cite{Cvetic:1999xp}. In this setting it was shown that in addition to anti-twist solutions \cite{Hosseini:2021fge,Boido:2021szx}, 
twist solutions \cite{Ferrero:2021etw} are also possible depending on the value of the magnetic fluxes and the deficit angles on the spindle.
In all of these examples, it is straightforward to calculate the central charge of the $d=2$ SCFT from the gravity solution. This can be compared with a field theory calculation that uses anomaly polynomials and $c$-extremisation \cite{Benini:2012cz}, and one finds exact agreement.

Similar investigations of SCFTs in $d=3,5, 6$ dimensions that are compactified on a spindle have also been made \cite{Ferrero:2020twa,Ferrero:2021wvk,Cassani:2021dwa,Ferrero:2021ovq,Couzens:2021rlk,Faedo:2021nub,Giri:2021xta,Couzens:2021cpk}. 
Furthermore, compactifying on higher-dimensional orbifolds is also possible
\cite{Cheung:2022ilc} (see also \cite{Suh:2022olh}): in particular, the case of the $\mathcal{N}=(0,2)$, $d=6$ SCFT arising on M5-branes and then reduced on four-dimensional orbifolds, including a spindle fibred over a spindle, was studied in \cite{Cheung:2022ilc}. 
The goal of the present paper is to report on an investigation of the $d=4$ Leigh-Strassler (LS) SCFT compactified on a spindle.

Recall that the LS fixed point is a strongly coupled $\mathcal{N}=1$, $d=4$ SCFT which was identified in \cite{Leigh:1995ep}.
It can be obtained as the IR end point of an RG flow that starts in the UV from $SU(N)$ $\mathcal{N}=4$ SYM theory with the addition
of a mass deformation for one of the three adjoint chiral superfields. The LS fixed point has $SU(2)\times U(1)_R$ global symmetry which is inherited from the $SU(4)$ $R$-symmetry of $\mathcal{N}=4$ SYM.
In the large $N$ limit the LS fixed point is holographically dual to an $AdS_5\times S^5_{LS}$ solution first found in \cite{Khavaev:1998fb,Pilch:2000ej}. 
Moreover, the holographic RG flow solution that starts from $AdS_5\times S^5$ in the UV, dual to 
the mass deformed $\mathcal{N}=4$ SYM, and then flows to $AdS_5\times S^5_{LS}$ in the IR was constructed in \cite{Freedman:1999gp}.
An analysis of the LS theory placed on $\mathbb{R}^{1,1}\times \Sigma_g$ with a standard topological twist, 
where $\Sigma_g$ is a Riemann surface of genus $g$, was made in \cite{Bobev:2014jva}. For genus $g>1$, 
by constructing $AdS_3\times \Sigma_g$ solutions of a sub-truncation $D=5$ maximal supergravity it was shown that the compactified LS theory flows to an $\mathcal{N}=(0,2)$, $d=2$ SCFT in the IR in the large $N$ limit. 

Here we consider the $d=4$ LS theory placed on $\mathbb{R}^{1,1}\times \Sigma$
where $\Sigma$ is a spindle with an azimuthal symmetry. Using the same sub-truncation of $D=5$ maximal gauged supergravity that was used in \cite{Bobev:2014jva}
we will construct an associated class of supersymmetric $AdS_3\times \Sigma$ solutions. After uplifting to type IIB  these give rise to a new class of $AdS_3\times Y_7$ solutions, with $Y_7$
a smooth manifold consisting of an $S^5$ fibration over $\Sigma$, that are dual to a new class of $\mathcal{N}=(0,2)$, $d=2$ SCFTs. 
A novel feature is that the $D=5$ gauged supergravity solution contains both neutral and charged scalar fields. Ensuring that the charged scalars are regular at the poles of the spindle requires a generalisation of the analysis of \cite{Ferrero:2021etw}.
While we have constructed some analytic $AdS_3\times \Sigma$ solutions to the BPS equations,
the generic solutions, all of which are in the anti-twist class, have been constructed numerically.
However, remarkably, we are able to show that the central charge can be expressed analytically in terms of the deficit angles of the poles and the magnetic flux through the spindle.
This enables us to make a comparison with a field theory calculation associated with the LS theory using anomaly polynomials and $c$-extremisation, and we find exact agreement.

The plan of the rest of the paper is as follows. In section \ref{sec:sugramodel} we introduce the $D=5$ supergravity model that we use to construct the new solutions. In section \ref{sec:ads3}
we present the $AdS_3\times \Sigma$ ansatz of interest and analyse the resulting BPS equations, which consist of a set of coupled ODEs. 
We identify conserved charges as well as
elucidate the boundary conditions that are required in order to obtain an $AdS_3\times \Sigma$ solution with appropriately quantised fluxes and regular scalar fields. This allows us to obtain an analytic expression for the central charge and also the fluxes in terms of the boundary conditions.
In section \ref{sec:results} we present the analytic solutions
as well as discuss the numerically constructed solutions. Section \ref{sec:fthy} carries out a field theory computation of the central charge.
We conclude with some discussion in section \ref{sec:disc}, including some
outlook on the possibility of constructing RG flow solutions that would connect with our new solutions.

We have five appendices. In appendix \ref{app:a} we discuss how the supergravity model arises from a truncation of maximal $D=5$ gauged supergravity, with some small differences in detail with regard to \cite{Khavaev:2000gb} and other papers. 
Our discussion  in the appendix is for a more general class of theories, as in \cite{Bobev:2010de}, that maintain three gauge fields and two neutral scalar fields, as in the STU model, plus four complex scalar fields. We also discuss a truncation to minimal $D=5$ gauged supergravity that is associated with the $AdS_5$ LS fixed point which allows us to construct the analytic $AdS_3\times \Sigma$ solutions.
In appendix \ref{susyvars} we derive the BPS equations for the $AdS_3$ ansatz of interest. We also show that the BPS equations can be
recast as supersymmetric $D=4$ Janus-like equations of the type discussed in a $D=5$ context in \cite{Arav:2020obl} which provides a helpful alternative perspective. In appendix \ref{app:c} we generalise the analysis of supersymmetric spindles
given in \cite{Ferrero:2021etw} to include charged complex scalar fields.
In appendix \ref{app:e} we analyse the possibility of having specific charged conformal Killing spinors on $\mathbb{R}^{1,1}\times\Sigma$, which would naturally arise on the boundary of putative RG flow solutions from $AdS_5$ to $AdS_3\times \Sigma$, finding that they only arise in the context of the standard topological twist.
In appendix \ref{app:d} we recall some features of the analytic $AdS_3\times\Sigma$ solutions of the STU model and discuss how RG flows from these fixed points to the new fixed points might be possible.

\section{The supergravity model}\label{sec:sugramodel}
We will use a $U(1)^3\subset SO(6)$ consistent truncation of maximal gauged supergravity in $D=5$ that 
keeps a metric, three gauge fields $A^{(1)},A^{(2)},A^{(3)}$, two real and neutral scalars $\alpha,\beta$ 
and a single complex scalar field $\zeta\equiv \varphi e^{i\theta}$ which is charged with respect to a specific linear combination of the three $U(1)$'s. This model was used in \cite{Bobev:2014jva} and can be obtained as a truncation of a more general class of models with four charged scalar fields that was 
presented in \cite{Khavaev:2000gb,Bobev:2010de} (see appendix \ref{app:a} for further discussion).
The bosonic part of the Lagrangian, in a \emph{mostly minus} signature, is given by 
\begin{align}\label{model1text}
\mathcal{L} =& -\tfrac{1}{4} R
 + \tfrac{1}{2}(\partial \varphi)^2+  \tfrac{1}{8}\sinh^2 2\varphi \left(D\theta \right)^2
 + 3 (\partial \alpha)^2 + (\partial \beta)^2   - \mathcal{P} \nn
&
 - \tfrac{1}{4}\left[ e^{4\alpha-4\beta} F^{(1)}_{\mu\nu}F^{(1)\mu\nu}
+ e^{4\alpha+4\beta} F^{(2)}_{\mu\nu}F^{(2)\mu\nu}
+ e^{-8\alpha} F^{(3)}_{\mu\nu}F^{(3)\mu\nu} \right]\nn
&+\tfrac{1}{2}\epsilon^{\mu\nu\rho\sigma\delta}F^{(1)}_{\mu\nu}F^{(2)}_{\rho\sigma}A^{(3)}_\delta\,,
\end{align} 
where
\begin{align}\label{dthetastext}
 D\theta&\equiv d \theta +g \left(A^{(1)} + A^{(2)} - A^{(3)} \right)\,.
\end{align}
The scalar potential $\mathcal{P}$ is given by
\begin{equation}\label{model2text}
\mathcal{P} = \frac{g^2}{8} \left[ \left( \frac{\partial W}{\partial \varphi} \right)^2 + \frac{1}{6}  \left( \frac{\partial W}{\partial \alpha} \right)^2 + \frac{1}{2} \left( \frac{\partial W}{\partial \beta} \right)^2 \right] - \frac{g^2}{3} W^2 \, ,
\end{equation}
where $W$ is the ``superpotential" defined by
\begin{align}\label{superpottext}
W = -\frac{1}{4} \Big[ 
 &( e^{-2\alpha-2\beta}+e^{-2\alpha+2\beta } - e^{ 4\alpha } ) \cosh2\varphi 
 + ( e^{-2\alpha-2\beta}+e^{-2\alpha+2\beta } +3 e^{ 4\alpha } )   \Big] \, .
\end{align}
Notice that the model has a $\mathbb{Z}_2$ symmetry
\begin{align}\label{z2symact}
\beta\to-\beta,\qquad A^{(1)}\leftrightarrow A^{(2)}\,.
\end{align}

The gravity theory \eqref{model1text}-\eqref{superpottext} is not supersymmetric but we can determine the conditions needed to be satisfied in order that a solution preserves some of the supersymmetry of the maximal gauged supergravity theory.
From the gravitino variations we require
\begin{align}\label{firstgravpass2text}
&\Big(\nabla_\mu -iQ_\mu-\frac{ig}{6}W\gamma_\mu
-\frac{1}{12}H_{\nu\rho}(\gamma^{\nu\rho}\gamma_\mu+2\gamma^\nu\delta^\rho_\mu)\Big)\epsilon=0\,,
\end{align}
where $\epsilon$ is a complex $D=5$ Dirac spinor, $\nabla_\mu=\partial_\mu+\tfrac{1}{4}\omega_{\mu ab}\gamma^{ab}$ and
\begin{align}\label{HQdefstext}
H_{\mu\nu} &\equiv  e^{2\alpha-2\beta} F^{(1)}_{\mu\nu} + e^{2\alpha + 2\beta} F^{(2)}_{\mu\nu} + e^{-4\alpha} F^{(3)}_{\mu\nu}  \, ,\nn
Q_\mu&\equiv -\frac{g}{2}(A^{(1)}_\mu+A^{(2)}_\mu+A^{(3)}_\mu)
-\frac{1}{4}(\cosh2\varphi-1)D_\mu\theta\,.
\end{align}
We highlight that the supersymmetry parameters are charged just with respect to the $R$-symmetry gauge field given by
\begin{align}\label{rgfielddeftext}
A^R_\mu \equiv -g(A^{(1)}_\mu+A^{(2)}_\mu+A^{(3)}_\mu)\,,
\end{align}
and have charge $1/2$. Also, notice that for vanishing complex scalar, $\varphi=0$, we have $A^R_\mu=2Q_\mu$.
Vanishing of the remaining supersymmetry variations is guaranteed if
\begin{align}\label{gauginovarstext}
[\gamma^\mu\partial_\mu \alpha +\frac{ig}{12}\partial_\alpha W-\frac{1}{12}(e^{2\alpha-2\beta} F^{(1)}_{\mu\nu} + e^{2\alpha+2\beta} F^{(2)}_{\mu\nu} - 2 e^{-4\alpha} F^{(3)}_{\mu\nu} )\gamma^{\mu\nu}]\epsilon&=0\,,\nn
{}[\gamma^\mu\partial_\mu \beta +\frac{ig}{4}\partial_\beta W-\frac{1}{4}(-e^{2\alpha-2\beta} F^{(1)}_{\mu\nu} +e^{2\alpha+2\beta} F^{(2)}_{\mu\nu})\gamma^{\mu\nu}]\epsilon&=0\,,\nn
{}[\gamma^\mu\partial_\mu \varphi +\frac{ig}{2}\partial_{\varphi}W+i\partial_{\varphi} Q_\mu\gamma^\mu]\epsilon&=0\,.
\end{align}

This gravity model admits the maximally supersymmetric $AdS_5$ vacuum solution with vanishing scalar fields and the $AdS_5$ metric having radius squared equal to $4/g^2$.
Within the associated dual $\mathcal{N}=4$ SYM theory we can identify the scalar fields $\alpha,\beta$ with bosonic mass deformations living in ${\bf 20}'$ of $SO(6)$ and $\zeta$ with fermionic mass deformations living in the ${\bf 10}$ of $SO(6)$.
If $X_a$ are the six real scalars and $\lambda$ one of the four fermions of $\mathcal{N}=4$ SYM theory then we have, schematically, 
\begin{align}\label{opfieldmapztext}
\Delta=2:\qquad \qquad  \alpha &\quad \leftrightarrow \quad \tr(X_1^2+X_2^2+X_3^2+X_4^2-X_5^2-X_6^2)\,,\nn
   \beta &\quad \leftrightarrow \quad \tr(X_1^2+X_2^2-X_3^2-X_4^2)\,,\nn
\Delta=3:\qquad \qquad  \zeta &\quad \leftrightarrow \quad \tr(\lambda\lambda+\text{cubic in $X_a$})\,,
\end{align}
where $\Delta$ is the conformal scaling dimension of the operator. 
It also admits\footnote{The model also admits an $SU(3)$ invariant $AdS_5$ solution which does not preserve supersymmetry and is known to be unstable \cite{Bobev:2010de}.}
a supersymmetric LS $AdS_5$ solution \cite{Khavaev:1998fb} with radius squared equal to $9/(2^{4/3}g^2)$ with
\begin{align}\label{lsfpsc}
e^{6\alpha}=2,\qquad e^{2\varphi}={3}\,,\qquad \beta=0\,,
\end{align}
and, of course, vanishing gauge fields. 
This solution preserves $SU(2)\times U(1)_R$ symmetry and, after uplifting to type IIB \cite{Pilch:2000ej},
are dual to the $d=4$, $\mathcal{N}=1$ LS SCFT \cite{Leigh:1995ep}. 
As is well known there is an RG flow between $\mathcal{N}=4$ SYM, deformed by the relevant fermion mass operator with $\Delta=3$ given in \eqref{opfieldmapztext}, and the corresponding holographic solution was found in \cite{Freedman:1999gp}.

It is also helpful to consider some further truncations of the gravity model.
If we set the charged scalar field to zero, $\zeta=0$, in \eqref{model1text} we obtain the STU model \cite{Cvetic:1999xp}, with two real scalars $\alpha,\beta$ and
three gauge-fields.
From the STU model, we can further truncate to minimal $D=5$ gauged supergravity by setting all gauge fields equal
$A^{(1)}=A^{(2)}=A^{(3)}$ as well as setting the real scalars to zero $\alpha=\beta=0$.
In particular, the $AdS_3\times \Sigma$ solutions of the STU model \cite{Ferrero:2021etw} (extending \cite{Hosseini:2021fge,Boido:2021szx})
and of minimal gauged supergravity \cite{Ferrero:2020laf} 
arise in this model, and can be uplifted on $S^5$ to obtain solutions of type IIB, as discussed in those papers.

There is also a different way to truncate to $D=5$ minimal gauged supergravity, associated with the LS fixed point. One sets the scalar fields
as in \eqref{lsfpsc} and also sets $A^{(1)}=A^{(2)}=\frac{1}{2}A^{(3)}$ (see appendix \ref{mingstrunc}). Thus, the 
$AdS_3\times \Sigma$ solutions of minimal gauged supergravity \cite{Ferrero:2020laf}  can be uplifted on $S^5$ in a different way to that discussed in
\cite{Ferrero:2020laf} and are dual to new $d=2$, $\mathcal{N}=(0,2)$ SCFTs; we shall discuss this further in section \ref{sec:resultsanalytic}.
We can also relax this truncation by just setting $A^{(1)}=A^{(2)}$ as well as $\beta=0$. This model preserves
$SU(2)\times U(1)_R$ symmetry. It contains the LS fixed point solution and also the RG flow solution that starts off with the vacuum 
$AdS_5$ in the UV and ends at the LS $AdS_5$ solution in the IR. One might have thought that this ``LS truncation" would be a good starting point to  construct additional $AdS_3\times \Sigma$ solutions but, as it turns out, there are no further solutions 
in this sector. Thus, we continue with
the larger truncation \eqref{model1text}-\eqref{superpottext} which preserves $U(1) \times U(1)_R$ symmetry, which we refer to as the \emph{extended LS truncation}.

\section{$AdS_3$ ansatz}\label{sec:ads3}
We are interested in constructing $AdS_3\times\Sigma$ solutions lying within the ansatz
\begin{align}\label{ads3ans}
ds^2&=e^{2V}ds^2(AdS_3)-(f^2dy^2+h^2 dz^2)\,,\nn
A^{(i)}&=a^{(i)}dz\,,
\end{align}
where $ds^2(AdS_3)$ is a unit radius metric on $AdS_3$ and $V,f,h,a^{(i)}$ are functions of $y$ only. Notice that we can use different gauge choices for $f$ by changing the $y$ coordinate
and later we will choose\footnote{If we dimensionally reduce on the $z$ direction
one gets a Janus-type ansatz in a conformal gauge  as discussed in appendix \ref{janusapp}.}
``conformal gauge"  with $f=e^V$.
We also assume that the scalar fields $\alpha,\beta,\varphi$ are functions of $y$ only.
To avoid PDEs, from the gauge equations of motion we must then take the phase of the complex scalar field, $\theta$, 
to be linear in $z$, $\theta=\bar\theta z$, with $\bar\theta$ a constant. In particular, we then have 
\begin{align}
Q_\mu dx^\mu\equiv Q_z dz\,,
\end{align}
with $Q_z =Q_z(y)$.

We are particularly interested in solutions where $\Sigma$, the space parametrised by $(y,z)$, is a compact spindle with an azimuthal symmetry generated by $\partial_z$. Compactness can be achieved by taking $y\in [y_1,y_2]$, with $h(y_i)=0$ and $z$ to be a periodic coordinate.
 The precise boundary conditions on the other fields at the poles of
the spindle, located at $y=y_i$ will be discussed below. We will be interested in such $AdS_3\times\Sigma$ solutions that preserve
supersymmetry and hence are dual to $d=2$ $\mathcal{N}=(0,2)$ SCFTs. The associated BPS equations will be summarised below.

We will utilise an orthonormal frame 
\begin{align}\label{orthframe}
e^a=e^V \bar e^a,\qquad e^3=f dy,\qquad e^4=hdz\,,
\end{align} 
where $\bar e^a$ is an orthonormal frame for $ds^2(AdS_3)$.
The frame components of the field strengths can then be written as
\begin{align}
F^{(i)}_{34}=f^{-1}h^{-1}(a^{(i)})'\,.
\end{align}

It will be very helpful to note that with this ansatz, two of the equations of motion for the gauge fields can be immediately integrated.
Explicitly, we find the gauge equations of motion are equivalent to
\begin{align}\label{gaugeintmot}
e^{3V}\left(e^{4\alpha-4\beta}F^{(1)}_{34}-e^{4\alpha+4\beta}F^{(2)}_{34}\right)&=\mathcal{E}_F\,,\nn
e^{3V}\left(e^{4\alpha-4\beta}F^{(1)}_{34}+e^{4\alpha+4\beta}F^{(2)}_{34}+2e^{-8\alpha}F^{(3)}_{34}\right)&=\mathcal{E}_R\,,\nn
(e^{3V-8\alpha}F^{(3)}_{34})'&=-e^{3V}fh^{-1}\frac{g}{4}\sinh^2 2\varphi D_z\theta \,,
\end{align}
where $\mathcal{E}_F$ and $\mathcal{E}_R$ are constants and $D_z\theta= (\bar\theta+ga^{(1)}+ga^{(2)}-ga^{(3)})$.

\subsection{BPS equations}\label{susyvarstext}
To analyse the Killing spinor equations\footnote{The analysis is somewhat similar to that of \cite{Bobev:2010de} who considered double wick rotated backgrounds
with the $AdS_3$ factor replaced by an $S^3$.}
we use the orthonormal frame given in \eqref{orthframe}.
We next decompose the Clifford algebra via
\begin{align}\label{eq:5d_clifford_deftext}
\gamma^m = & \Gamma^m \otimes \sigma^3, \qquad  
\gamma^3 =  {1} \otimes i  \sigma^1, \qquad
\gamma^4 =   {1} \otimes  i  \sigma^2, 
\end{align}
with $\Gamma^m=(\sigma^2,  i\sigma^3, i\sigma^1)$ gamma matrices in $D=3$.
We write the Killing spinor as
\begin{align}
\epsilon=\psi\otimes\chi\,,
\end{align}
with $\psi$ a two component spinor on $AdS_3$ which satisfies
\begin{align}
D_m\psi=\frac{i}{2}\kappa\Gamma_m \psi\,,
\end{align}
with $\kappa=\pm 1$ determining the chirality of the supersymmetry of the
dual $d=2$ SCFT i.e. $\mathcal{N}=(0,2)$ or $(2,0)$.
 After some analysis, summarised in appendix \ref{susyvars}, we find that the two component spinor $\chi$ can be
written in the form
\begin{align}\label{chiralityspin}
\chi  =e^{V/2}e^{is z}\begin{pmatrix}\sin\frac{\xi}{2} \\ \cos\frac{\xi}{2} \end{pmatrix}\,,
\end{align}
where the constant $s$ is the gauge-dependent charge of the spinor under the action of the azimuthal Killing vector $\partial_z$. 
Notice, for later use, that at points where
$\xi=0$ the spinor $\chi$ has negative chirality with respect to $\sigma^3$, while when 
$\xi=\pi$ the spinor $\chi$ has positive chirality. 

The solutions of interest to us have $\sin\xi$ not identically equal to zero. Then, as we show in the appendix \ref{susyvars}, for points with $\sin\xi\ne 0$ the 
BPS equations associated with these Killing spinors can be written in the form
\begin{align}\label{summbbpsapptext}
f^{-1}\xi'&=gW\cos\xi+2\kappa e^{-V}\,,\nn
f^{-1} V' &= \frac{g}{3} W \sin\xi\,,\nn
f^{-1}\alpha'&=-\frac{g}{12}\partial_\alpha W\sin\xi\,,\nn
f^{-1}\beta'&=-\frac{g}{4}\partial_\beta W\sin\xi\,,\nn
f^{-1}\varphi'&=-\frac{g}{2}\frac{\partial_{\varphi}W}{\sin\xi}\,,\nn
f^{-1}\frac{h'}{h}&=\frac{1}{\sin\xi}\Big(2\kappa e^{-V}\cos\xi+\frac{gW}{3}(1+2\cos^2\xi)\Big)\,,
\end{align}
along with the two constraint equations
\begin{align}\label{summconstext}
(s-Q_z)\sin\xi&=-\frac{1}{2}g  Wh\cos\xi-\kappa h e^{-V}\,,\nn
\frac{g}{2}\partial_{\varphi}W\cos\xi&=\partial_{\varphi} Q_z \sin\xi h^{-1}\,.
\end{align}
Furthermore the field strength components in the orthonormal frame are given by
\begin{align}\label{fstrengthsbpstext}
e^{2\alpha-2\beta}F^{(1)}_{34}&=-\frac{g}{12}[4W-\partial_\alpha W+3\partial_\beta W]\cos\xi-\kappa e^{-V}\,,\nn
e^{2\alpha+2\beta}F^{(2)}_{34}
&=-\frac{g}{12}[4W-\partial_\alpha W-3\partial_\beta W]\cos\xi-\kappa e^{-V}\,,\nn
e^{-4\alpha}F^{(3)}_{34}
&=-\frac{g}{6}[2W+\partial_\alpha W]\cos\xi-\kappa e^{-V}\,.
\end{align}

\subsection{Integrals of motion}
An important observation is that we can integrate a combination of the BPS equations. Specifically,
by calculating the derivative of $he^{-V}$ we deduce that
\begin{align}\label{hemvszeqtext}
he^{-V}=k \sin\xi\,,
\end{align}
where $k$ is a constant.
This shows that at points where $h$ vanishes, which 
will correspond to the poles of the spindle in the solutions of interest, $\sin\xi$ also vanishes. It is also helpful to notice that from \eqref{summbbpsapptext}, \eqref{summconstext}
we can then also write the equation for $\xi'$ as
\begin{align}\label{xiderivnice}
\xi'&=-2k^{-1}(s-Q_z)(e^{-V}f)\,,
\end{align}
while the two constraints \eqref{summconstext} can now be written in the form
\begin{align}\label{summconstext2}
(s-Q_z)&=-k[\frac{1}{2}g  We^V\cos\xi+\kappa]\,,\nn
\frac{g}{2}\partial_{\varphi}W\cos\xi&=k^{-1}e^{-V}\partial_{\varphi} Q_z\,.
\end{align}

We can also use the expressions for the field strengths \eqref{fstrengthsbpstext}
to rewrite the two integrals of motion \eqref{gaugeintmot}, arising from the gauge field equations of motion, to obtain
\begin{align}\label{conschges}
\mathcal{E}_R&=e^{2V}[2ge^V \cos\xi-2\kappa(e^{-4\alpha}+e^{2\alpha}\cosh 2\beta)]\,,\nn
\mathcal{E}_F&=2\kappa e^{2V}e^{2\alpha}\sinh 2\beta\,.
\end{align}

\subsection{Boundary conditions for spindle solutions}\label{bcsspinsols}
It is convenient to now work in conformal gauge with
\begin{align}\label{confgauge}
f=e^V\,,
\end{align} so that the metric takes the form
\begin{align}
ds^2&=e^{2V}[ds^2(AdS_3)-ds^2_\Sigma]\,,
\end{align}
with
\begin{align}
ds^2_\Sigma=dy^2+k^2\sin^2\xi dz^2\,,
\end{align}
and the constant $k$ defined in \eqref{hemvszeqtext}.

We are interested in constructing supersymmetric solutions where $ds^2_\Sigma$ is a metric on a spindle with an azimuthal symmetry, which is specified by two relatively prime integers $n_N,n_S\ge 1$.
The poles are
taken to lie at $y=y_{N,S}$ and with deficit angles $2\pi(1-\frac{1}{n_{N,S}})$, respectively, and $z$ is a periodic coordinate with period $\Delta z$ which we fix to be 
\begin{align}
\Delta z=2\pi\,.
\end{align}
In order to ensure that the gauge fields $gA^{(i)}$ are connections on $U(1)$ orbibundles over $\Sigma$ we need to impose that the magnetic fluxes through the spindle are suitably quantised: 
\begin{align}\label{genfluxqc}
\frac{1}{2\pi}\int_\Sigma gF^{(i)}=\frac{p_i}{n_N n_S}\,,\qquad p_i\in\mathbb{Z}\,,
\end{align}
as discussed in \cite{Ferrero:2021etw} and also summarised in appendix \ref{app:c}. This normalisation is fixed by how the $D=5$ solution
is uplifted on $S^5$ to give a type IIB solution\footnote{From the uplifting formula (2.1) of \cite{Cvetic:1999xp} we see that
$[gA^{i}]_{them}$ is a canonically normalised $U(1)$ connection since their $\phi_i$ are periodic coordinates on $S^5$ with periods  
$\Delta\phi_i=2\pi$. By comparing their (2.9) with
our \eqref{model1text} we conclude that $g_{us}=2g_{them}$, $A^{(i)}_{us}=1/2 A^{i}_{them}$ and hence 
$[gA^{(i)}]_{us}$ is canonically normalised.} 
and ensures that we will obtain an $AdS_3\times Y_7$ solution with $Y_7$ a smooth manifold consisting of an $S^5$ bundle over the spindle $\Sigma$ \cite{Ferrero:2021etw}.

A novel feature in the present set up is the presence of the complex scalar field
$\zeta$ which is charged under a particular linear combination of the three $U(1)$'s. We analyse the boundary conditions that need to be imposed on such scalars in appendix \ref{app:c}, extending the analysis of \cite{Ferrero:2021etw}.
We show there that if $\zeta$ is non-vanishing at a pole then we must have the gauge invariant condition $D\theta=0$ at that pole:
\begin{align}\label{textdthetazero}
\varphi|_{N,S}\ne 0 \quad \Rightarrow\quad D\theta|_{N,S}=0\,,
\end{align} 
respectively.
Furthermore, in the case that
$\zeta$ is non-vanishing at \emph{both} poles, which is the case that we shall study\footnote{It would be interesting to determine
whether or not there are spindle solutions where the complex scalar vanishes at just one of the poles.},
the associated $U(1)$ flux through the spindle must vanish:
\begin{align}\label{nobrokenflux}
\frac{1}{2\pi}\int_\Sigma g(F^{(1)}+F^{(2)}-F^{(3)})&=\frac{1}{2\pi}\int_\Sigma d(D\theta)\nn
&=0\,.
\end{align}
Given that the $R$-symmetry flux is quantised, as we recall in a moment, we just need to impose one more condition to ensure that
the general flux quantisation conditions \eqref{genfluxqc} are satisfied for all gauge-fields. 

We now consider the fermions. The coupling of the fermions to the $R$-symmetry gauge field $A^R$, defined in
\eqref{rgfielddeftext}, is exactly the same as in \cite{Ferrero:2021etw}. There are then
two cases, the twist and the anti-twist, which are specified by the $R$-symmetry flux: 
\begin{align}\label{rsymflux}
\frac{1}{2\pi}\int_\Sigma F^R\equiv\frac{1}{2\pi}\int_\Sigma -g(F^{(1)}+F^{(2)}+F^{(3)})&=\pm\frac{n_N+n_S}{n_N n_S}\,,\qquad \text{Twist}\,,\nn
&=\pm\frac{n_S-n_N}{n_N n_S}\,,\qquad \text{Anti-twist}\,.
\end{align}
The $\pm$ signs refer to the chirality of the spinors at the poles, as we make more precise below, and we recall that for the twist class, the spinors have the same chirality at the two poles while
for the anti-twist class they have opposite chirality.

The behaviour of the $R$-symmetry gauge field and the azimuthal charge of the spinor at the poles, which depends on the choice of gauge, was discussed in
\cite{Ferrero:2021etw}. From \eqref{HQdefstext} we note that at the poles of the spindle the $R$-symmetry gauge field
is equal to $2Q_\mu$; this is obviously true if the complex scalar vanishes at the pole, but it is also true if
it doesn't since, as noted above, in this case  we demand that $D\theta=0$ at the pole. Recall, that $s$ is the azimuthal charge of the Killing spinor, in a given gauge. Then from (2.36), (2.37) of  \cite{Ferrero:2021etw} (see also the discussion in appendix \ref{app:c})
we can conclude that the behaviour of the gauge-invariant quantity $s-Q_z$ at the poles can be taken to satisfy
\begin{align}\label{sminqnpoletext}
(s-Q_z)|_N=\pm\frac{1}{2n_N},\qquad (s-Q_z)|_S&=\mp \frac{1}{2n_S},\quad\text{Twist}\,, \nn
(s-Q_z)|_S&=\pm  \frac{1}{2n_S},\quad\text{Anti-twist}\,.
\end{align}

With these general comments in mind, we will analyse the BPS equations in more detail.
In doing so we will recover some of these results directly. However, remarkably, 
we will be able to achieve significantly more, almost completely fixing the behaviour of all fields at the poles.
Furthermore, we will be able to obtain an analytic expression for
the central charge of the dual field theory in terms of the spindle data $(n_N,n_S)$ 
for both the twist and the anti-twist class. Consistency of the resulting conditions will in fact eliminate the possibility of any twist solutions, leaving us just with the possibility of anti-twist solutions. The existence of such
solutions can be demonstrated numerically, as we discuss in section \ref{sec:results}.

\subsubsection{Analysis of the BPS equations}

We begin by noting that at the poles of the spindle
we have $k\sin\xi\to 0 $ and hence, taking $k\ne 0$ we have $\cos\xi\to \pm 1$. We write 
$\cos\xi_{N,S}= (-1)^{t_{N,S}}$, with $t_{N,S}\in \{0,1\}$.
The poles are located at $y=y_{N,S}$ and we label them so that $y_N<y_S$ and $y\in[y_N,y_S]$. By assumption, the deficit angles at the poles are
$2\pi(1-\frac{1}{n_{N,S}})$, with $n_{N,S}\ge 1$, and hence from the metric we should demand that $|(k\sin\xi)'|_{N,S}=\frac{1}{n_{N,S}}$. 
It is convenient to use the symmetry \eqref{sym1} to fix 
\begin{align}\label{symassumph}
h\ge 0\,,\qquad \Leftrightarrow\qquad k\sin\xi\ge 0\,,
\end{align} 
using \eqref{hemvszeqtext}.
We then must have $(k\sin\xi)'|_N>0$ and $(k\sin\xi)'|_S<0$. 
With $y\in [y_N,y_S]$ we therefore
impose
\begin{align}
(k\sin\xi)'|_{N,S}=\frac{(-1)^{l_{N,S}}}{n_{N,S}},\qquad l_N=0,l_S=1\,.
\end{align}

From the general analysis of \cite{Ferrero:2021etw} we know that there are two classes to consider, the twist and the anti-twist. In the twist class the preserved spinors have the same chirality at the two poles
while in the anti-twist class they have opposite chirality. Thus, we have\footnote{\label{footsym3}In the anti-twist case we could utilise the symmetry \eqref{sym3} and a relabelling of the poles to reduce to either $(t_N,t_S)=(1,0)$ or $(t_N,t_S)=(0,1)$, but to simplify the presentation we don't do that.}
\begin{align}\label{coszetapole}
\cos\xi|_{N,S}= (-1)^{t_{N,S}}\,;\qquad &\text{Twist:} \qquad\quad \,\,(t_N,t_S)=(1,1)\quad\, \text{or} \quad(0,0) ,\nn
&\text{Anti-Twist:} \quad \,\,(t_N,t_S)=(1,0)\quad \text{or} \quad(0,1) \,.
\end{align}

Next, from the BPS equation \eqref{xiderivnice} we have $(k\sin\xi)'=-2\cos\xi(s-Q_z)$ and hence we can write
\begin{align}\label{qresults}
(s-Q_z)|_{N,S}=\frac{1}{2n_{N,S}}(-1)^{l_{N,S}+t_{N,S}+1}\,,
\end{align}
exactly as in \eqref{sminqnpoletext}.
We can now obtain an expression for the $R$-symmetry flux. From 
\eqref{HQdefstext} we have $\tfrac{1}{2}F^R=dQ+d(\frac{1}{4}(\cosh2\varphi-1)D\theta)$. Given our ansatz and integrating over the spindle,
the second term on the right hand side will not contribute since either $\varphi=0$ at a pole or if not then $D_z\theta=0$ at that pole as in \eqref{textdthetazero} (see appendix \ref{app:c}). The contribution from
the first term can be evaluated using \eqref{qresults} and we obtain
\begin{align}\label{rsymflux2}
\frac{1}{2\pi}\int_\Sigma F^R\equiv\frac{1}{2\pi}\int_\Sigma -g(F^{(1)}+F^{(2)}+F^{(3)})
&= \frac{n_N (-1)^{t_S+1} + n_S(-1)^{t_N+1}}{n_N n_S} \, ,
\end{align}
exactly as in \eqref{rsymflux}.

Continuing, we next note that $\partial_{\varphi}Q_z= -\frac{1}{2}\sinh 2\varphi D_z\theta$ and hence using the same argument
as in the previous paragraph, we deduce that $\partial_{\varphi}Q_z=0$ at the poles. From the BPS constraint \eqref{summconstext2} we then
deduce $\partial_{\varphi}W=0$ at the poles. Thus,
\begin{align}\label{betbc}
\partial_{\varphi}Q_z|_{N,S}=\partial_{\varphi}W|_{N,S}=0.
\end{align}
It is interesting to point out that we can reach this conclusion another way: from the BPS equation for $\varphi$ in \eqref{summbbpsapptext}, we see that whenever $\sin\xi=0$ we require $\partial_{\varphi}W$ to also vanish to ensure
that $\varphi$ stays finite. Thus, at a pole we deduce $\partial_{\varphi}W|_{N,S}=0$ and hence from \eqref{summconstext2} 
$\partial_{\varphi}Q_z|_{N,S}=0$ also.

To make further progress we will now assume that the complex scalar is non-vanishing at both poles which implies (see appendix \ref{app:c})
\begin{align}\label{nonvancsc}
\varphi|_{N},\varphi|_{S}\ne 0\,,
\qquad\Rightarrow \qquad D_z\theta|_{N}= D_z\theta|_{S}= 0\,.
\end{align}
This immediately implies that the flux of the $U(1)$ which the complex scalar is charged under must vanish, as in \eqref{nobrokenflux}:
\begin{align}
\frac{1}{2\pi}\int_\Sigma g(F^{(1)}+F^{(2)}-F^{(3)})&= (D_z\theta)|^{y_S}_{y_N}=0\,.
\end{align}
With the $R$-symmetry flux quantised as in \eqref{rsymflux2}, we just need to impose one more condition to ensure that
the general flux quantisation condition \eqref{genfluxqc} is satisfied for all gauge-fields.

Proceeding, given \eqref{nonvancsc}, the second condition in \eqref{betbc} and the expression for $W$ in \eqref{superpottext}
imply that 
\begin{align}\label{betbc2}
(e^{6\alpha}-2\cosh 2\beta)|_{N,S}=0\,,\quad\Rightarrow \quad
W|_{N,S}=-e^{4\alpha}|_{N,S}\,.
\end{align}
We now want to examine the value of the conserved charges $\mathcal{E}_R, \mathcal{E}_F$, given in \eqref{conschges}, at both poles.
It is convenient to first define two quantities 
\begin{align}\label{emmbc}
M_{(1)}\equiv g e^{4\alpha} e^V,\qquad M_{(2)} \equiv -2\kappa + 2 M_{(1)} \cos\xi\,,
\end{align}
and for future reference we note that $M_{(1)}>0$.
 We then notice
from \eqref{conschges} that we can write 
\begin{align}\label{eesemmms}
\mathcal{E}_R &= \frac{M_{(1)}^2}{g^2} \left[ -\kappa + M_{(2)} e^{-12\alpha} \right]+
\kappa e^{2V+2\alpha}(e^{6\alpha}-2\cosh 2\beta)\, ,\nn
(\mathcal{E}_F)^2 &= \frac{M_{(1)}^4}{g^4} \left[ 1 - 4e^{-12\alpha} \right]
+\frac{M_{(1)}^4}{g^4}e^{-12\alpha} \left[ 4\cosh^2 2\beta-e^{12\alpha} \right]
\,,
\end{align}
and observe that the second term on the right hand side vanishes at the poles, for both lines, as a consequence of
\eqref{betbc2}. Furthermore, using \eqref{betbc2} and
the first constraint equation in \eqref{summconstext2} we deduce that at the poles
\begin{align}\label{emmbcpoles}
M_{(1)}|_{N,S}&= 2 (-1)^{t_{N,S}} \kappa - \frac{1}{kn_{N,S}} (-1)^{l_{N,S}}\,,\nn
M_{(2)}|_{N,S}&= 2\kappa - \frac{2}{k n_{N,S}} (-1)^{l_{N,S}+t_{N,S}} \,.
\end{align}
We can now evaluate $\mathcal{E}_R$ at each of the two poles and, being constant, 
these values must be equal to each other. The same applies to $ (\mathcal{E}_F)^2 $ and
so we deduce that $\alpha_{N,S}$ are fixed by solving a set of linear equations:
\begin{equation}\label{emmlinsys}
\begin{pmatrix}
-4 M_{(1)}^4|_N & 4 M_{(1)}^4|_S \\
M_{(1)}^2 |_NM_{(2)}|_N & - M_{(1)}^2|_S M_{(2)}|_S 
\end{pmatrix}
\begin{pmatrix}
e^{-12\alpha_{N}} \\
e^{-12\alpha_{S}}
\end{pmatrix}
=
\begin{pmatrix}
M_{(1)}^4|_S - M_{(1)}^4|_N \\
-\kappa M_{(1)}^2|_S + \kappa M_{(1)}^2|_N
\end{pmatrix}\,.
\end{equation}

We now take stock of these results. For a given $\kappa=\pm1 $, consider fixing spindle data $n_N, n_S, t_{N,S}$,
along with the constant $k$. Solving \eqref{emmlinsys} then allows us to obtain $\alpha$ at both poles of the spindle, $\alpha_{N,S}$, in 
terms of $(n_N, n_S, t_{N,S},k)$. 
We can then also obtain\footnote{\label{signbetaftn}Notice that the sign of $\mathcal{E}_F$ is the sign of $\kappa\beta$. Therefore, for any solution for 
$\alpha_{N,S}$, one gets two possible boundary conditions of $\beta$ for a given $\kappa$.} $\beta$ and $V$ at both poles using \eqref{betbc2} and the definition of $M_{(1)}$ in
\eqref{emmbc}. Notice from \eqref{eesemmms} we must have 
\begin{align}\label{alphconst}
0 < e^{-12\alpha_{N,S}}\leq 1/4\,,
\end{align} 
and this restricts the allowed range 
of $k$ for given spindle data $n_N, n_S, t_{N,S}$. We also highlight that from \eqref{emmbcpoles}, the values of $\alpha,\beta$ and $V$
at the poles just depend\footnote{The two-forms $F_{34}e^3\wedge e^4$ evaluated at the poles have an  
additional explicit dependence on $k$.}
on the combinations $kn_{N}$ and $kn_S$.

Thus, for each $\kappa=\pm 1$, given spindle data $n_N, n_S, t_{N,S}$ and the constant $k$ satisfying the above constraint, we have specified the values of $\alpha,\beta,V$
at both poles. We also have $\xi$ at both poles from \eqref{coszetapole}. We therefore just need to specify the value of $\varphi$ at one
of the poles, and recall we have assumed that this is non-vanishing, in order to obtain a solution to the BPS equations. We will construct such solutions numerically in section \ref{sec:results}. Before doing that we will show that further progress can be made by analysing the fluxes.

\subsubsection{Fluxes}
We now show, remarkably, that we can obtain expressions for the overall fluxes of the three gauge fields on the spindle in terms of 
the spindle data at the poles as expressed in the previous subsection in terms of $n_N, n_S, t_{N,S},k$. Furthermore, we will be able to invert these and obtain an expression for $k$
in terms of $n_N, n_S, t_{N,S}$ and the quantised flavour flux $p_F\equiv p_1-p_2$, where $p_i$ are defined in \eqref{genfluxqc}.

We first note that we can use the BPS equations \eqref{summbbpsapptext} to re-express the field strengths \eqref{fstrengthsbpstext} as total $y$ derivatives of expressions that depend only on the scalar fields, warp factors, the angle $\xi$ and $k$:
\begin{equation}
F^{(i)}_{yz} = (a^{(i)})' = (\mathcal{I}^{(i)})' \, ,
\end{equation}
where
\begin{align}\label{eq:IntegratedFluxesExpr1}
& \mathcal{I}^{(1)} \equiv \frac{1}{2} k e^V \cos\xi \, e^{-2\alpha+2\beta} \, , \quad
\mathcal{I}^{(2)} \equiv \frac{1}{2} k e^V \cos\xi \, e^{-2\alpha-2\beta} \, , \quad
 \mathcal{I}^{(3)} \equiv \frac{1}{2} k e^V \cos\xi \, e^{4\alpha} \, .
\end{align}
In appendix \ref{sec:b1} we comment on how we obtained these expressions.
This immediately allows us to express the fluxes in terms of pole data:
\begin{equation}
\frac{p_i}{n_N n_S} \equiv 
\frac{1}{2\pi} \int_\Sigma g F^{(i)} = g \left. \mathcal{I}^{(i)} \right|^S_N\,.
\end{equation}

On the other hand, we can use the expressions from the previous subsection 
to relate these expressions at the poles to $n_N, n_S, t_{N,S},k$ as follows:
\begin{align}\label{iidefs}
& \mathcal{I}^{(1)} |_{N,S}= \frac{1}{2} ( \mathcal{I}_0 \pm \mathcal{I}_\Delta )|_{N,S} \, , \nn
& \mathcal{I}^{(2)} |_{N,S}= \frac{1}{2} ( \mathcal{I}_0 \mp \mathcal{I}_\Delta ) |_{N,S}\, , \nn
& \mathcal{I}^{(3)} |_{N,S}= \mathcal{I}_0 |_{N,S}\, , 
\end{align}
where
\begin{align}\label{eq:IntegratedFluxesI0}
&\mathcal{I}_0|_{N,S} \equiv \frac{1}{2g} k M_{(1)}|_{N,S} (-1)^{t_{N,S}}  \, ,\nn
&\mathcal{I}_\Delta|_{N,S} \equiv \frac{1}{2g} k M_{(1)}|_{N,S} (-1)^{t_{N,S}} \sqrt{1-4e^{-12\alpha_{N,S}}} \, ,
\end{align}
and the $\pm$ sign in \eqref{iidefs} depends on the chosen sign of $\beta$ (see footnote \ref{signbetaftn} and we also note that we could fix this sign using the symmetry \eqref{z2symact} if desired).
Notice that $\mathcal{I}_0|^S_N$ is independent of $k$, and in fact
\begin{align}\label{fluxioexp}
g \mathcal{I}_0 |^S_N =\frac{1}{2} (\mathcal{I}^{(1)} +\mathcal{I}^{(2)}+\mathcal{I}^{(3)})|^S_N &=\frac{1}{2} \frac{n_N (-1)^{t_S} + n_S(-1)^{t_N}}{n_N n_S} \, ,\nn
(\mathcal{I}^{(1)} +\mathcal{I}^{(2)}-\mathcal{I}^{(3)})|_{N,S}&=0\,.
\end{align}
in agreement with the overall $R$-symmetry flux and vanishing of the flux of the broken $U(1)$, respectively.
Furthermore,
proper quantization conditions for all three fluxes then imply the following:
\begin{enumerate}
\item $ p_3 = n_N n_S\, g \mathcal{I}_0 |^S_N \in \mathbb{Z} $. This condition depends only $n_N, n_S$, and is satisfied as long as $ n_N (-1)^{t_S} + n_S(-1)^{t_N}\, $ is even.
\item $ p_F \equiv p_1 - p_2 = \text{sign}(\beta) n_N n_S\, g \mathcal{I}_\Delta |^S_N \in \mathbb{Z} $. This condition translates to a condition on $k$.
\item Notice that $p_3=2p_1-p_F$ and hence $ p_3, p_F $ are both even or both odd.
\end{enumerate}

We can now invert the above relations to obtain $k$ in terms of $n_N, n_S, t_{N,S}$ and $p_F$. To this end, we first note that from \eqref{eq:IntegratedFluxesI0} we have
\begin{equation}
\mathcal{I}_\Delta |_{N,S} = \frac{g}{2} k \frac{(-1)^{t_{N,S}}}{M_{(1)} |_{N,S}} |\mathcal{E}_F | \, ,
\end{equation}
where $\mathcal{E}_F$ is the conserved charge given in \eqref{eesemmms}. 
Using the fact that $\mathcal{E}_F$ is conserved and in particular the same value at the two poles, we have
\begin{equation}
g \mathcal{I}_\Delta |^S_N = \frac{g^2}{2} k |\mathcal{E}_F| \left[ \frac{(-1)^{t_S}}{M_{(1)}|_S} - \frac{(-1)^{t_N}}{M_{(1)}|_N} \right] =
g^2 |\mathcal{E}_F| \frac{(-1)^{t_N+t_S+1}}{M_{(1)}|_S M_{(1)}|_N} \left(   g\mathcal{I}_0 |^S_N \right) \, ,
\end{equation}
and thus
\begin{equation}
\frac{p_F^2}{p_3^2} = \frac{(M_{(1)}|_N)^2}{(M_{(1)}|_S)^2} (1 - 4 e^{-12\alpha_N}) \, .
\end{equation}
We also note that $\text{sign}(\beta)=\text{sign}(p_F)\text{sign}(p_3)(-1)^{t_N+t_S+1}$.
Using the previous expressions for $M_{(1)}$ and $e^{-12\alpha}$ at the poles expressed in terms of $n_N, n_S, t_S$ and $k$ obtained by solving \eqref{emmlinsys}, we can invert this expression and find an expression for $k$. In the twist class, labelled by $t_N$ as in \eqref{coszetapole}, 
we find
\begin{equation}\label{eq:IntegratedFluxeskExprTwist}
k = {\kappa}(-1)^{1+t_N} \frac{(n_N+n_S)^2 (n_N^2 - n_N n_S + n_S^2) - 4 n_N n_S p_F^2}{n_N n_S (n_N - n_S) \left( 3(n_N + n_S)^2 + 4 p_F^2 \right)} \, ,\quad\text{Twist}\,,
\end{equation}
while for the anti-twist class, labelled by $t_N$ as in \eqref{coszetapole},  we get
\begin{equation}\label{eq:IntegratedFluxeskExprAntiTwist}
k ={\kappa}(-1)^{t_N} \frac{(n_N-n_S)^2 (n_N^2 + n_N n_S + n_S^2) + 4 n_N n_S p_F^2}{n_N n_S (n_N + n_S) \left( 3 (n_N - n_S)^2 + 4 p_F^2 \right)} \, ,
\quad\text{Anti-Twist}\,.
\end{equation}

\subsubsection{Central charge}\label{cchgecalc}
We now obtain an expression for the central charge of the dual $d=2$, $\mathcal{N}=(0,2)$ SCFT.
We first note that the five-dimensional Newton's constant is given by $(G_{(5)})^{-1}=N^2g^3/(4\pi)$. This is associated with
the $AdS_5\times S^5$ vacuum solution, dual to $\mathcal{N}=4$ SYM with gauge group $SU(N)$, 
having an $a$-central charge $a^{\mathcal{N}=4}=\pi R^3/8G_{(5)}=N^2/4$, where recall that the radius of the $AdS_5$ space for this solution is $R=2/g$.
Similarly the $LS$ $AdS_5$ solution with radius $R^{LS}=3/(2^{2/3}g)$ gives an $a$-central charge $a^{LS}=27N^2/128$.
The three-dimensional Newton's constant for a theory admitting a unit radius $AdS_3$ solution is then 
$(G_{(3)})^{-1}=(G_{(5)})^{-1}\Delta z\int^{y_S}_{y_N} e^V |f h| dy$ (not using conformal gauge \eqref{confgauge} here) and the 
$d=2$ central charge is given by $c=(3/2)(G_{(3)})^{-1}$.

We next note the remarkable result that the integrand appearing in the central charge can again be expressed as a total derivative,\begin{align}
e^V f h=ke^{2V}f\sin\xi=-\frac{k}{2\kappa}\left(e^{3V}\cos\xi\right)'\,,
\end{align}
and hence the central charge can be expressed in terms of data at the poles. Recall that we have been working in conformal gauge \eqref{confgauge} so that the integrand appearing in the central charge integral is $e^{2V}|h|$. Recall, 
furthermore, that we used a symmetry of the BPS equations to set $h\ge 0$ in
\eqref{symassumph} and so we can remove the absolute value in the integrand. We thus obtain
\begin{align}
c&=6N^2\left(\frac{g}{2}\right)^3(-\frac{k}{2\kappa})[e^{3V}\cos\xi]^S_N\nn
&=-\frac{3N^2 k}{8\kappa}(M^3_{(1)}|_Se^{-12\alpha_S}(-1)^{t_S}-M^3_{(1)}|_Ne^{-12\alpha_N}(-1)^{t_N})\,,
\end{align}
where we used $\Delta z =2\pi$. 
Using the results for $k$ and $M_{(1)}$ in terms of $n_N, n_S, t_S$ and $p_F$ from the previous subsections, we can now get expressions for the central charge in terms of these parameters. 

For the twist case we find
\begin{equation}\label{twistcentchgegrav}
c =  \kappa(-1)^{t_N+1}\frac{3(n_S + n_N) \left[ (n_N+n_S)^2 - 4 p_F^2 \right] \left[ 3 (n_N + n_S)^2 + 4 p_F^2 \right]}{32n_N n_S \left[ (n_N+n_S)^2 (n_N^2 - n_N n_S + n_S^2) - 4 n_N n_S p_F^2  \right]}N^2\, .
\end{equation} 
By construction $c>0$. Recall that we also have $M_{(1)}|_{N,S}>0$ and in addition that
$ 0 < e^{-12\alpha_{N,S}} \leq 1/4$ from \eqref{alphconst}.  Remarkably we find that these inequalities eliminate the twist case completely! 

On the other hand for the anti-twist case we find
\begin{equation}\label{cchgeat}
c = \kappa(-1)^{t_N}\frac{3 (n_N - n_S) \left[ (n_N-n_S)^2 - 4 p_F^2 \right] \left[ 3 (n_N - n_S)^2 + 4 p_F^2 \right]}{32n_N n_S \left[ (n_N-n_S)^2 (n_N^2 + n_N n_S + n_S^2) + 4 n_N n_S p_F^2  \right]}N^2 \, .
\end{equation} 
Imposing exactly the same positivity conditions we just mentioned, imposes the
following constraints for the anti-twist class:
\begin{align}\label{attslambdaf}
t_N=0,\kappa>0,\quad\text{or}\quad t_N=1,\kappa<0 \qquad \Rightarrow\quad (n_N-n_S)>2|p_F|\ge 0\,,\nn
t_N=0,\kappa<0,\quad\text{or}\quad t_N=1,\kappa>0 \qquad \Rightarrow\quad (n_S-n_N)>2|p_F|\ge 0\,.
\end{align}
We also recall from the conclusions listed just below \eqref{fluxioexp} that in order for $p_3$ to be an integer we require $n_N-n_S$ to be even and $p_3,p_F$ are both even or both odd. 
The individual fluxes for these anti-twist solutions
can be expressed in the form
\begin{align}\label{indivflux}
p_1&=(-1)^{t_N}\frac{n_S-n_N}{4n_N n_S}+\frac{p_F}{2}\,,\nn
p_2&=(-1)^{t_N}\frac{n_S-n_N}{4n_N n_S}-\frac{p_F}{2}\,,\nn
p_3&=(-1)^{t_N}\frac{n_S-n_N}{2n_N n_S}\,.
\end{align}

We find it remarkable that we have been able to obtain these results without solving the BPS equations. Our numerical investigations indicate that providing the above conditions are satisfied
then a spindle solution to the BPS equations in the anti-twist class with properly quantised flux does in fact always exist.

For the special case that we set $p_F=0$, we obtain
\begin{equation}\label{cchgeatsimp}
c = \kappa(-1)^{t_N}\frac{4 (n_N - n_S)^3 }{3n_N n_S  (n_N^2 + n_N n_S + n_S^2) }a^{LS} \, ,
\end{equation} 
where $a^{LS}=27N^2/128$ is the $a$-central charge of the $d=4$ LS SCFT. We also recall from point 3 below \eqref{fluxioexp}
that for this case $n_N - n_S$ must be divisible by 4 to ensure that we have properly quantised fluxes.
In fact for the $p_F=0$ case there is an associated class of analytic $AdS_3\times\Sigma$ solutions to the BPS equations, as
we discuss in the next section.
Notice that the central charge \eqref{cchgeatsimp} for these solutions has the same form
as other type IIB solutions obtained by uplifting solutions of minimal gauged supergravity on Sasaki-Einstein spaces \cite{Ferrero:2020laf}.

\section{Solving the BPS equations}\label{sec:results}
\subsection{Analytic solutions for $p_F=0$}\label{sec:resultsanalytic}
When $p_F=0$ we can find analytic $AdS_3\times \Sigma$ solutions in the anti-twist class
by utilising the truncation to minimal $D=5$ gauged supergravity
that is associated with the LS $AdS_5$ solution (see appendix \ref{mingstrunc}) and the class of $AdS_3\times \Sigma$ solutions given in
\cite{Ferrero:2020laf}. Note that these solutions are not given in the conformal gauge \eqref{confgauge}. 

Specifically we find that the following solves the general BPS equations given in 
\eqref{summbbpsapptext}-\eqref{summconstext} along with \eqref{fstrengthsbpstext}. 
We set the scalars to have the same values as in the LS $AdS_5$ vacuum solution,
\begin{align}\label{lsfpsc2}
e^{6\alpha}=2,\qquad e^{2\varphi}={3}\,,\qquad \beta=0\,,
\end{align}
and set $\theta=0$. The metric and gauge fields are given by
\begin{align}\label{minlssol}
ds^2&=\frac{9}{2^{4/3} g^2}\left[ \frac{4y}{9}ds^2(AdS_3)-\frac{y}{q(y)} dy^2-\frac{q(y)}{36 y^2}c_0^2 dz^2\right]\,,\nn
A^{(1)}=A^{(2)}=\tfrac{1}{2} A^{(3)}&=-\left[\frac{c_0\kappa}{8g}\left(1-\frac{a}{y}\right)+\frac{ s}{2g}\right]dz\,,
\end{align}
and we also have
\begin{align}
\sin\xi=-\frac{\sqrt{q(y)}}{2y^{3/2}},\qquad
\cos\xi=\kappa\frac{3y-a}{2 y^{3/2}}\,.
\end{align}
The function $q(y)$ is the cubic
\begin{align}
q(y)=4y^3-9y^2+6 ay-a^2\,,
\end{align}
where the constant $a$ and the constant $c_0$ given in \eqref{minlssol} are given by
\begin{align}
a&=\frac{ (n_S-n_N)^2(2n_S+n_N)^2(n_S+2n_N)^2     }{4(n_S^2+n_S n_N+n_N^2)^3}\,,\nn
c_0&=\frac{2(n_S^2+n_S n_N+n_N^2)}{3n_S n_N(n_S+n_N)}\,.
\end{align}
We take $n_S>n_N$ with $y\in [y_N,y_S]$ where $y_N,y_S$ are the two smallest roots of the cubic $q(y)$ given by
\begin{align}
y_N&=\frac{(n_S^2+n_S n_N -2n_N^2)^2}{4(n_S^2+n_S n_N+n_N^2)}\,,\qquad
y_S=\frac{(n_N^2+n_S n_N -2n_S^2)^2}{4(n_S^2+n_S n_N+n_N^2)}\,.
\end{align}

The central charge of the dual $d=2$ SCFT can be calculated directly from this solution
and we precisely recover \eqref{cchgeatsimp} for the case $(-1)^{t_N+1}\kappa=+1$.

\subsection{Numerical solutions for $p_F\ne 0$}
We now use the results of section \ref{bcsspinsols} to numerically construct $AdS_3\times \Sigma$ solutions to the BPS equations in the anti-twist class when
$p_F\ne 0$.
In section \eqref{bcsspinsols}, which we recall used the conformal gauge \eqref{confgauge},
 we showed how given spindle data $n_N, n_S, t_S$ and for each $\kappa=\pm 1$, we can obtain
the values of $\alpha,\beta,V$ and $\xi$ at both poles of the spindle. 
Furthermore, for a specified value of the flux $p_F$ 
the integration constant $k$ is fixed via \eqref{eq:IntegratedFluxeskExprAntiTwist}. 
Thus, we can search an $AdS_3\times \Sigma$ solution by specifying the value of $\varphi$ at one of the poles, integrating the BPS equations and then looking for a solution for which $\sin\xi$ vanishes for some other finite value of the coordinate $y$. 

More explicitly, we start the integration at $y=y_N$, say, and for convenience we take $y_N=0$. At this pole we have
$\sin\xi=0$ by assumption. For generic values of $\varphi$ at $y=0$ we find that solving the BPS equations with a given $k$ and initial values for all functions as described above leads to solutions which do not have another finite value of $y=y_S$ for which $\sin\xi=0$ and hence will not give rise to a solution with $\Sigma$ a compact spindle (in fact the solutions have divergent $\varphi$ at some finite value of $y$). We therefore need to scan over a range of initial values for $\varphi$ in order to find the compact spindle solutions. When we do find such a solution then our procedure automatically guarantees that the fluxes are all suitably quantised. We have carried
out this procedure for about 100 different values\footnote{Note that the symmetry \eqref{z2symact} can be used to flip the sign of $p_F$.} of $n_N,n_S,p_F$ and provided that the condition given in
\eqref{attslambdaf} is satisfied we have always found an explicit (and unique) spindle solution. This provides strong evidence that the condition 
\eqref{attslambdaf} is sufficient for the anti-twist solutions to exist.

\begin{figure}[h!]
\begin{center}
\includegraphics[scale=.7]{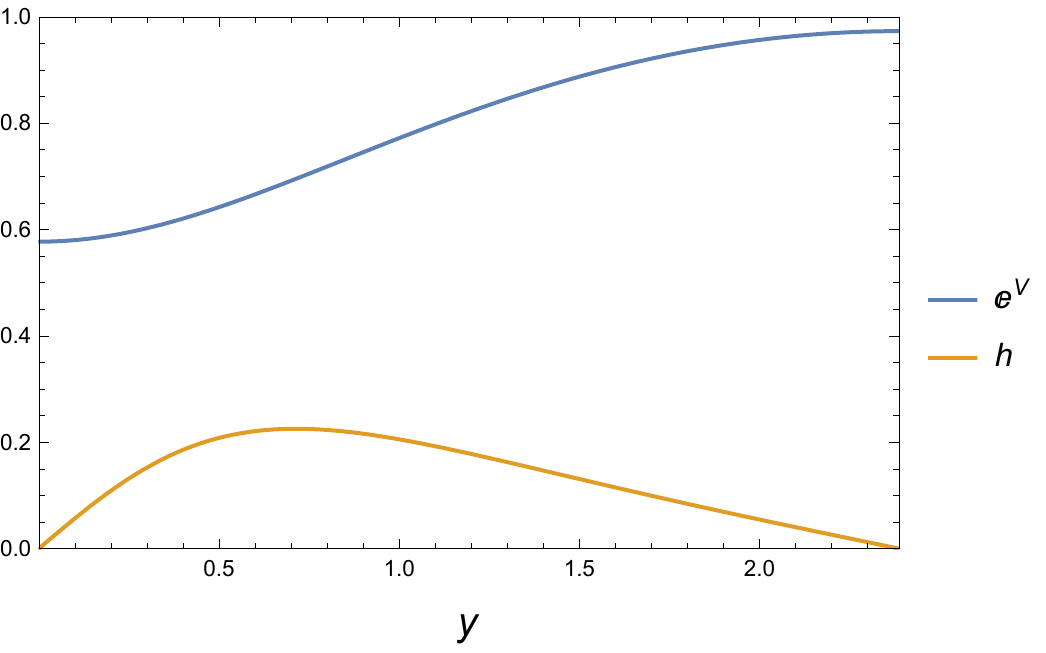}
\includegraphics[scale=.7]{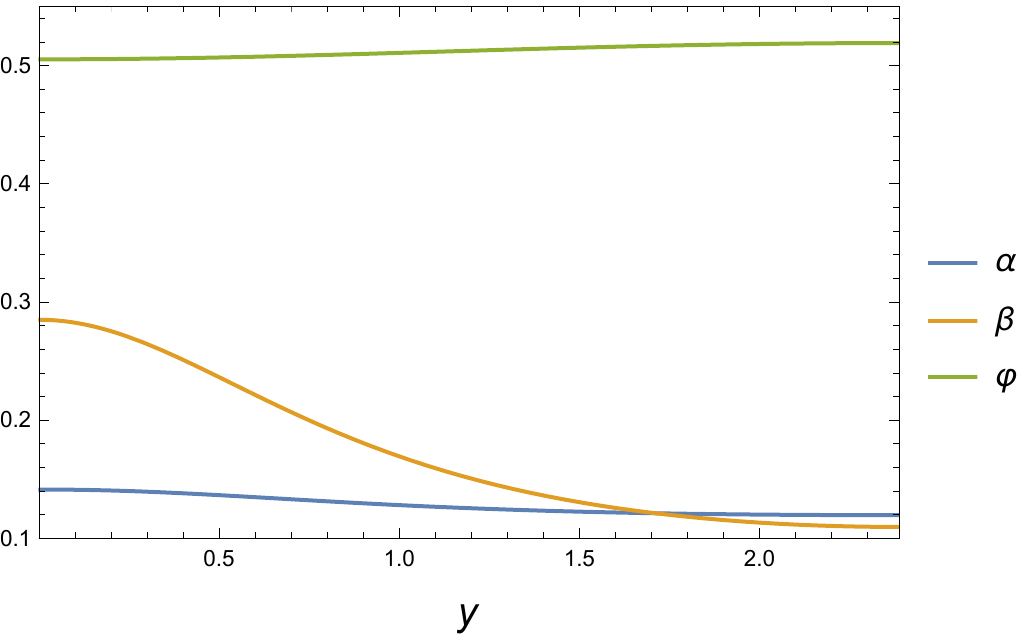}
\caption{An $AdS_3\times \Sigma$ solution in the anti-twist class with $n_N=1$, $n_S=7$ and $p_F = 1$. In the left panel we have
plotted the metric functions $e^V$ and $h$ while in the right panel we have plotted the three scalar functions $\alpha,\beta$ and $\varphi$.}
\label{fig:example_wavefns}
\end{center}
\end{figure}

We have used the numerically constructed $AdS_3\times \Sigma$ solutions to directly calculate the central charge
by carrying out the integral discussed in section \ref{cchgecalc}. Our numerical solutions are such that we find agreement with
the analytic result \eqref{cchgeat} to the numerical accuracy that we used, roughly of order $10^{-8}$.
In figure \ref{fig:example_wavefns} we have plotted the metric and scalar functions for a representative example with 
$n_N=1$, $n_S=7$ and $p_F=\pm 1$. For this example we found the explicit value of the scalar field at the poles
are given by $\varphi|_N\sim 0.50516$ and  $\varphi|_S\sim 0.51895$.

In table \ref{table1} we have presented the values of $\varphi$ at the poles for a few more illustrative solutions.
Recall that we require $n_S-n_N$ to be even and that $p_F=0$ solutions exist when $n_S-n_N$ is divisible by four.
The individual fluxes are given in \eqref{indivflux}.
\begin{table}[h!]
\footnotesize
\begin{center}
\begin{tabular}{|c|l|}
\hline
$(n_N,n_S)$ & $\qquad\qquad\qquad\quad$ Value of $(\varphi_N$, $\varphi_S)$ \\\hline
(1,5)&$p_F=0$: ($\tfrac{1}{2}\ln 3$, $\tfrac{1}{2}\ln 3$);\\\hline
(1,7)&$|p_F|=1$: (0.50516, 0.51895);\\\hline
(1,9)&$p_F=0$: ($\tfrac{1}{2}\ln 3$, $\tfrac{1}{2}\ln 3$);\quad $|p_F|=2$: (0.44938, 0.47913);\\\hline
(1,11)&$|p_F|=1$: (0.53187, 0.54025);\quad $|p_F|= 3$: (0.40487, 0.44471);\\\hline
(1,13)&$p_F=0$: ($\tfrac{1}{2}\ln 3$, $\tfrac{1}{2}\ln 3$);\quad $|p_F|= 2$: (0.50093, 0.52388);\\
&\qquad\qquad\qquad\qquad\qquad\quad$|p_F|= 4$: (0.37019, 0.41593);\\\hline
(3,7)&$p_F=0$: ($\tfrac{1}{2}\ln 3$, $\tfrac{1}{2}\ln 3$);\\\hline
(3,11)&$p_F=0$: ($\tfrac{1}{2}\ln 3$, $\tfrac{1}{2}\ln 3$);\quad $|p_F|= 2$: (0.45936, 0.46794);\\\hline
(3,13)&$|p_F|=1$: (0.53441, 0.53747);\quad $|p_F|= 3$: (0.41749, 0.43002);\\\hline
(5,9)&$p_F=0$: ($\tfrac{1}{2}\ln 3$, $\tfrac{1}{2}\ln 3$);\\\hline
(5,11)&$|p_F|= 1$: (0.51118, 0.51257);\\\hline
(5,13)&$p_F=0$: ($\tfrac{1}{2}\ln 3$, $\tfrac{1}{2}\ln 3$);\quad $|p_F|= 2$: (0.46175, 0.46547);\\\hline
\end{tabular}
\caption{Values of the scalar field $\varphi$ at the north and south pole, $(\varphi_N$, $\varphi_S$), for some representative
$AdS_3\times \Sigma$ solutions for spindle data $n_N,n_S$ and flavour flux $p_F$. In general we have
$n_S-n_N$ even. The $p_F=0$ solutions exist when $n_S-n_N$ is divisible by four; these are the analytic solutions of section \ref{sec:resultsanalytic}.\label{table1}}
\end{center}
\end{table}

\section{Field theory analysis}\label{sec:fthy}
In this section we analyse the $\mathcal{N}=1$ $d=4$ LS SCFT compactified on a spindle. Assuming that the resulting theory flows to an $\mathcal{N}=(0,2)$ $d=2$ SCFT at low energies, we can calculate the central charge using anomaly polynomials and $c$-extremisation. The calculation runs along similar lines to \cite{Ferrero:2020laf}.

The LS SCFT has $U(1)_R\times SU(2)$ symmetry. We want to focus on an abelian
subgroup of the flavour symmetry, 
$U(1)_F\subset SU(2)$. From table 2 of \cite{Bobev:2014jva}, in the large $N$ limit 
we have
\begin{align}
Tr(R^3)&=\frac{3N^2}{4}\,,\qquad
Tr(R F^2)=-\frac{N^2}{4}\,,\qquad
Tr(R^2 F)=Tr(F^3)=0\,,
\end{align}
where the trace is over chiral fermions. We can thus write the 6-form anomaly polynomial for the LS theory, in the large $N$ limit, as
\begin{align}
\mathcal{A}_{LS}=\frac{N^2}{8}[c_1(R)^3 -c_1(R) c_1(F)^2]\,,
\end{align}
where $c_1(R)$ and $c_1(F)$ are the first Chern classes for the $U(1)_R$ and $U(1)_F$ bundles.

We now consider the LS theory compactified on a spindle $\Sigma$ with an azimuthal symmetry, 
specified by relatively prime integers $n_N,n_S\ge 1$, 
and background magnetic fluxes for both $U(1)_R$, consistent with supersymmetry, and $U(1)_F$. We want to carry out $c$-extremisation 
 \cite{Benini:2012cz} for the resulting $d=2$ theory allowing for a mixing of the $U(1)_R$, $U(1)_F$ and $U(1)_J$ symmetries, where $J$ generates azimuthal rotations on
the spindle. 

We let $y,z$ parametrise the spindle with $y\in [y_N,y_S]$ and $\Delta z=2\pi$.
Following \cite{Ferrero:2020laf} we introduce connection one-forms
\begin{align}
\mathscr{A}^{(R)} &=  \rho_R(y)\left(d z+\mathcal{A}_{\mathcal{J}}\right)\,,\qquad
\mathscr{A}^{(F)} =  \rho_F(y)\left(d z+\mathcal{A}_{\mathcal{J}}\right)\,,
\end{align} 
with curvatures $\mathscr{F}^{(R)} =  d\mathscr{A}^{(R)}$, $\mathscr{F}^{(F)} =  d\mathscr{A}^{(F)}$ where
\begin{align}
\mathscr{F}^{(R)} &=  \rho'_R(y)d y \wedge \left(dz+\mathcal{A}_{\mathcal{J}}\right)+\rho_R(y)\,\mathcal{F}_{\mathcal{J}}\,,\nn
\mathscr{F}^{(F)} &=  \rho'_F(y)d y \wedge \left(dz+\mathcal{A}_{\mathcal{J}}\right)+\rho_F(y)\,\mathcal{F}_{\mathcal{J}}\,,
\end{align}
and $\mathcal{F}_{\mathcal{J}}=d \mathcal{A}_{\mathcal{J}}$.
We have the flux conditions
\begin{align}\label{rhodiffs}
\frac{1}{2\pi}\int_\Sigma \mathscr{F}^{(R)}  &=  [\rho_R]^{y_S}_{y_N} =  \frac{p_R}{n_N n_S}\,,\nn
\frac{1}{2\pi}\int_\Sigma \mathscr{F}^{(F)}  &=  [\rho_F]^{y_S}_{y_N} =  \frac{p_F}{n_N n_S}\,,
\end{align}
with $p_R, p_F\in \mathbb{Z}$.

There are two possibilities, the twist and the anti-twist \cite{Ferrero:2021etw}. We will work in a gauge for the $R$-symmetry in 
which the spinors on the spindle are independent of the $z$ coordinate and we can write
\begin{align}\label{rhoR}
\rho_R(y_N)=\frac{(-1)^{t_N}}{n_N},\quad\rho_R(y_S)=\frac{(-1)^{t_S+1}}{n_N}\,.
\end{align}
with $p_R= \frac{(-1)^{t_S+1}}{n_N}+\frac{(-1)^{t_N+1}}{n_S}$. We have
$t_N=(0,1)$ and $t_S= t_N$ for the twist case and $t_S=t_N+1$ for the anti-twist case.
From the expression for the
flavour symmetry flux we can write 
\begin{align}
\rho_F(y_N)=\alpha_0,\quad\rho_F(y_S)=\frac{p_F}{n_N n_S}+\alpha_0\,,
\end{align}
where $\alpha_0$ is arbitrary; we will se that $\alpha_0$ will drop out of the final expressions for the central charge.

The curvature forms $\mathscr{F}^{(R)},\mathscr{F}^{(F)}$ define corresponding first Chern classes
$c_1(\mathcal{L}_R) = [\mathscr{F}^{(R)}/2\pi]$, $c_1(\mathcal{L}_F) = [\mathscr{F}^{(F)}/2\pi]$,
and similarly we define 
$c_1(\mathcal{J})=[\mathcal{F}_\mathcal{J}/2\pi]$. Following \cite{Ferrero:2020laf}, 
to obtain the $d=2$ anomaly polynomial $\mathcal{A}_{2\diff}$ we make the following substitution
in the $d=4$ anomaly polynomial $\mathcal{A}_{4\diff}$ :
\begin{align}
c_1(R)&\to c_1(R)+\frac{1}{2}c_1(\mathcal{L}_R)\,,\qquad
c_1(F)\to c_1(F)+c_1(\mathcal{L}_F)\,,
\end{align}
and then  integrate over $\Sigma$. Here the factor of 1/2 arises because\footnote{If the supercharge has charge 1 with respect to the gauge
field $A^R$, then it has charge 1/2 with respect to $2A^R$.} the field theory $R$-symmetry generator is normalised
so that the supersymmetry generator has charge 1, whereas the expressions in \eqref{rhoR} are for when it has charge 1/2, as in our earlier supergravity calculation. Performing the integral we find\footnote{Note that we can rewrite this expression as a ```gluing formula" analogous to section 4.1 of \cite{Hosseini:2021fge}.}
\begin{align}\label{anomaly2d}
\mathcal{A}_{2\diff}=  \int_{\Sigma}\mathcal{A}_{4\diff}\,
 = 
&\frac{N^2}{8}\Big\{
 c_1(R)^2(\frac{3}{2}[\rho_R]^{y_S}_{y_N})
 -c_1(F)^2( \frac{1}{2}[\rho_R]^{y_S}_{y_N})
\nn
& +c_1(\mathcal{J})^2(\frac{1}{8}[\rho_R^3]^{y_S}_{y_N}- \frac{1}{2} [\rho_R\rho_F^2]^{y_S}_{y_N} )
-c_1(R)c_1(F)(2 [\rho_F]^{y_S}_{y_N})
 \nn
 &+c_1(R)c_1(\mathcal{J})(\frac{3}{4}[\rho_R^2]^{y_S}_{y_N}- [\rho_F^2]^{y_S}_{y_N} )
 -c_1(F)c_1(\mathcal{J})( [\rho_R\rho_F]^{y_S}_{y_N})
\Big\}\,.
 \end{align}

Assuming that the LS theory compactified on a spindle flows to a SCFT in the IR we can determine 
the $d=2$ superconformal $R$-symmetry and central charge via $c$-extremisation \cite{Benini:2012cz}. Specifically,
the $d=2$ superconformal $R$-symmetry extremises the trial function
\begin{align}\label{centralfromtrace}
c_{\text{trial}}\, = \, 3\,\text{tr}\,\gamma_3 R^2_{\text{trial}}\,,
\end{align}
where
\begin{align}
R_{\text{trial}}\, = \, R+x F+\varepsilon\,\mathcal{J}\,,
\end{align}
with $x$ and $\varepsilon$ parametrising the mixing with the flavour symmetries.
Now the coefficient of $\frac{1}{2}c_1(L_i)c_1(L_j)$ in $\mathcal{A}_{2\diff}$ is $Tr\gamma^3 Q_i Q_j$ where
the global symmetry $Q_i$ is associated to the $U(1)$ bundle $L_i$ and $\gamma_3$ is the $d=2$ chirality operator.
From \eqref{anomaly2d}  we therefore have
\begin{align}
c_{\text{trial}}   =  
\frac{3N^2}{4}\Big[&
\frac{3}{2}[\rho_R]^{y_S}_{y_N}
-x^2(\frac{1}{2}[\rho_R]^{y_S}_{y_N})
+\varepsilon^2(\frac{1}{8}[\rho_R^3]^{y_S}_{y_N}-\frac{1}{2}  [\rho_R\rho_F^2]^{y_S}_{y_N} )\nn
&-x(2 [\rho_F]^{y_S}_{y_N})
+\varepsilon(\frac{3}{4}[\rho_R^2]^{y_S}_{y_N}- [\rho_F^2]^{y_S}_{y_N})
-x \varepsilon([\rho_R\rho_F]^{y_S}_{y_N})
\Big]\, .
\end{align}

We can now extremise with respect to $x,\varepsilon$ and then get a prediction for the central charge. 
First consider the twist case with $t_S=t_N$. We find the on-shell central charge is given by
\begin{equation}\label{cctwistcase}
c = (-1)^{t_N+1}\frac{3(n_S + n_N) \left[ (n_N+n_S)^2 - 4 p_F^2 \right] \left[ 3 (n_N + n_S)^2 + 4 p_F^2 \right]}{32n_N n_S \left[ (n_N+n_S)^2 (n_N^2 - n_N n_S + n_S^2) - 4 n_N n_S p_F^2  \right]}N^2\, .
\end{equation} 
This extremum occurs at
\begin{align}
\varepsilon_*&=(-1)^{t_N}\frac{(n_N-n_S)n_S n_N(  3(n_S+n_N)^2+4p_F^2    )}
{[(n_S+n_N)^2(n_S^2-n_S n_N+n_N^2)-4n_S n_N p_F^2]}\,, \nn
x_*&=(-1)^{t_N+1}\frac{(n_S+n_N)^2(n_N-2n_S)p_F+4n_Np_F^3}{(n_S+n_N)^2(n_S^2-n_S n_N+n_N^2)-4n_S n_N p_F^2}-\alpha_0\varepsilon_*\,.
\end{align}
Notice that $\alpha_0$ has dropped out of the expression of the on-shell central charge. Furthermore
note that there is a preferred value for $\alpha_0$ for which $x_*=0$.
Remarkably, the result for the central charge for the twist case given in \eqref{cctwistcase} is 
exactly the same as the gravity result \eqref{twistcentchgegrav} with $\kappa=+1$ (which has been implicitly assumed in the
field theory computation). However, we recall that there are, in fact, no supergravity solutions for the twist case.

Also, still for the twist case, notice that when we set $n_S=n_N=1$ we find that at the extremal point we have
\begin{align}\label{bobevpilchcase}
c&=(-1)^{t_N+1}\frac{3N^2}{4}(3+{p_F^2}),\quad \varepsilon_*=0,\quad x_*={p_F}\,.
\end{align}
We can now compare with the calculation for the case of a topological twist on an $S^2$ as discussed in
\cite{Bobev:2014jva}. Setting their genus $g=0$, if we identify our flavour $x$ mixing parameter with their $\epsilon$ and
our $p_F/2$ with their $\mathfrak{b}$ (which they note is half integer valued for $g=0$ just below their (2.12)), then we get exact agreement
for the on-shell central charge\footnote{We also note that if we set $\varepsilon=0$ at the start of the above calculation along with $n_N=n_S=1$ then we would find
an off-shell central charge $c_{trial}=(-1)^{t_N+1}\frac{3N^2}{4}(3+2(-1)^{t_N}{p_F}x-x^2)$, which similarly aligns with the result in 
\cite{Bobev:2014jva}.}, up to an overall sign. In fact we must have $t_N=+1$ for \eqref{bobevpilchcase} to be positive.

We now consider the anti-twist case with $t_S=t_N+1$.
We find the on-shell central charge is given by
\begin{equation}\label{cchgeatapp}
c = (-1)^{t_N}\frac{3 (n_N - n_S) \left[ (n_N-n_S)^2 - 4 p_F^2 \right] \left[ 3 (n_N - n_S)^2 + 4 p_F^2 \right]}{32n_N n_S \left[ (n_N-n_S)^2 (n_N^2 + n_N n_S + n_S^2) + 4 n_N n_S p_F^2  \right]}N^2 \, .
\end{equation} 
This result agrees exactly with the gravity calculation \eqref{cchgeat} $\kappa=+1$, and we recall that supergravity solutions do
exist for the anti-twist case.
This extremum occurs at
\begin{align}
\varepsilon_*&=(-1)^{t_N+1}\frac{(n_N+n_S)n_S n_N(  3(n_S-n_N)^2+4p_F^2    )}
{(n_N-n_S)^2(n_S^2+n_S n_N+n_N^2)+4n_S n_N p_F^2}\,, \nn
x_*&=(-1)^{t_N}\frac{(n_N-n_S)^2(n_N+2n_S)p_F+4n_N p_F^3}{(n_N-n_S)^2(n_S^2+n_S n_N+n_N^2)+4n_S n_N p_F^2)}-\alpha_0\varepsilon_*\,.
\end{align}
Note in the special case when we set $p_F=0$, we get the on-shell results
\begin{align}
c&=(-1)^{t_N}\frac{9(n_N-n_S)^3}{32n_S n_N (n_S^2+n_S n_N +n_N^2)}N^2\,,\nn
\varepsilon_*&=(-1)^{t_N+1}\frac{3(n_S+n_N)n_S n_N}{ n_S^2+n_S n_N +n_N^2}\,,\nn
x_*&=-\alpha_0\varepsilon_*\,.
\end{align}
in alignment with \eqref{cchgeatsimp}.

\section{Discussion}\label{sec:disc}
We have constructed a new class of supersymmetric $AdS_3\times\Sigma$ solutions of the extended LS model, a sub-truncation of $D=5$ maximal gauged supergravity. 
All of the solutions lie within the anti-twist class.
After uplifting to $D=10$ these give rise to supersymmetric $AdS_3\times Y_7$ solutions of type IIB supergravity with $Y_7$ a compact manifold that consists of an $S^5$ fibration over the spindle $\Sigma$. The fact that the orbifold singularities of the spindle disappear in the 
uplift to $Y_7$ is a consequence of the fact that the $U(1)^3$ fluxes in the $D=5$ solution have been suitably quantised \cite{Ferrero:2020laf,Ferrero:2020twa,Ferrero:2021etw}. 

A remarkable aspect of our analysis of the BPS equations is that we derived analytic expressions for the overall fluxes of the three gauge fields on the spindle in terms of the spindle data at the poles as well as an integration constant $k$. Furthermore, this enabled us to obtain
an analytic expression for the central charge of the dual SCFT expressed in terms of the deficit angles of the poles and the flavour magnetic flux through the spindle. It would be desirable to have a better understanding as to why this was possible, as it seems likely that it
will also apply in the context of some other supersymmetric solutions of supergravity theories.

The $AdS_3\times Y_7$ solutions are dual 
to a new class of $\mathcal{N}=(0,2)$, $d=2$ SCFTs and we have calculated the associated central charge in the
large $N$ limit. We have also made a direct comparison with a field theory computation. Specifically, we considered
the LS $d=4$ SCFT compactified on a spindle and, assuming that the resulting theory flows to an $\mathcal{N}=(0,2)$ SCFT
in the IR, we computed the central charge using anomaly polynomials and $c$-extremisation. Remarkably we find exact agreement with the gravity computation.

It is curious that all of the new $AdS_3\times\Sigma$ solutions lie within the anti-twist class. This is despite the fact that the associated
field theory computation for the central charge does not seem to rule out the twist case. This situation is 
somewhat  parallel to the analysis of the LS theory compactified on a Riemann surface of genus $g$, $\Sigma_g$ \cite{Bobev:2014jva}.
It was shown in \cite{Bobev:2014jva} that $c$-extremisation does not obviously rule out the genus $g=0$ or $g=1$ cases and yet 
$AdS_3\times \Sigma_g$ solutions were only found for $g>1$. It seems likely that these results are not unrelated.

It would be very interesting to construct black string solutions of type IIB supergravity that started off at the $AdS_5\times S^5_{LS}$ solution in the UV and ended up at the new $AdS_3\times Y_7$ solutions in the IR. 
The spindle horizon indicates that such black strings should be accelerating.
However, there are obstacles in constructing such solutions in a straightforward way. One might hope for solutions within gauged supergravity that preserve supersymmetry along the flow as
well as preserving the azimuthal symmetry of the spindle and the flavour symmetry preserved by the IR solution. 
Correspondingly the conformal boundary should be  of the form $\mathbb{R}^{1,1}\times \Sigma$ and preserve these symmetries. 
One can consider the simplest class of $AdS_3\times \Sigma$ solutions with vanishing flavour flux, $p_F=0$, which have been constructed analytically using minimal $D=5$ gauged supergravity \cite{Ferrero:2020laf}. For this class no black string solutions of minimal gauged supergravity are known to exist.
If they do exist, then the conformal boundary of such black string solutions should 
admit $d=4$ conformal Killing spinors charged under the $R$-symmetry\cite{Cheng:2005wk,Klare:2012gn,Cassani:2012ri}. However, in appendix \ref{app:e} we show that such Killing spinors only
lie within the twist class (in fact the usual topological twist) and not the anti-twist class. Perhaps black string solutions can be constructed 
within a bigger truncation of gauged supergravity or in type IIB supergravity itself, allowing for more general deformations on
the boundary and relaxing one or more of the conditions of supersymmetry, azimuthal and flavour symmetry.

Similar issues have been encountered for the class of supersymmetric $AdS_2\times \Sigma$ solutions of $D=4$ minimal gauged supergravity in the anti-twist class \cite{Ferrero:2020twa}. There it was shown 
that a general class of magnetically charged and accelerating black hole solutions in $AdS_4$ do in fact exist which
approach $AdS_2\times \Sigma$ solutions in the near horizon limit. However, the solutions that approach the supersymmetric locus have the peculiar feature that the conformal boundary gets pierced by an acceleration horizon and degenerates into
two pieces, each realising supersymmetry via a topological twist, but a different one on each component. It was also shown
that for a more general class of accelerating black hole solutions with the addition of
electric charge and rotation, solutions to the conformal Killing spinor equation do exist 
\cite{Ferrero:2020twa} and furthermore there are additional potential connections with a locus of complex solutions as discussed in \cite{Cassani:2021dwa}. It is clearly desirable to have a better understanding of anti-twist black hole and black string solutions.

In a complementary direction, it is also interesting to ask if there are RG flows that connect the new $AdS_3\times \Sigma$ solutions of the extended LS model with the analytic $AdS_3\times \Sigma$ solutions of the STU model \cite{Hosseini:2021fge,Boido:2021szx,Ferrero:2021etw}. For this to be possible
 we should demand that the fluxes through the spindle are the same for both solutions. This implies that we should consider the sub-class of STU models which have vanishing flux
for the $U(1)$ that is carried by the complex scalar in the extended LS model.
In appendix \ref{app:d} we show that imposing this condition on the solutions of the STU model
eliminates the twist class, in alignment with the fact that
we don't find any twist solutions of the extended LS model. On the other hand one finds that
there is a family of anti-twist solutions in the STU model with the same restrictions on the spindle data and fluxes
as in the extended LS model solutions. Furthermore, we find that central charge of these STU model solutions is always bigger than the central charge of the extended LS model solutions.
This strongly suggests that there should be supersymmetric RG flows that start off with the STU solutions in the UV and end up at the extended LS solutions in the IR.

A final comment is that it would be interesting to further analyse the geometry of $Y_7$ in the uplifted $AdS_3\times Y_7$ solution.
There has been significant recent progress in understanding the GK geometry \cite{Kim:2005ez,Gauntlett:2007ts}
on $AdS_3\times Y_7$ solutions of type IIB supergravity, dual to $\mathcal{N}=(0,2)$ $d=2$ SCFTS,
 which only have non-vanishing five-form flux, starting with \cite{Couzens:2018wnk}. The new $AdS_3\times Y_7$ solutions will have additional fluxes switched on and it would be very interesting to cast the geometry in the language of \cite{Couzens:2019iog}. Indeed this exercise could lead to the construction of additional new classes of solutions.

\section*{Acknowledgements}

\noindent 
We thank Seyed Morteza Hosseini for helpful discussions.
This work was supported in part by STFC grant ST/T000791/1
and by the H2020-MSCA-IF project RGxGRAV no. 101028617.
JPG is supported as a Visiting Fellow at the Perimeter Institute.

\appendix

\section{Gauged supergravity truncations}\label{app:a}
We use conventions consistent with those of maximal $SO(6)$ $D=5$ gauged supergravity in \cite{Gunaydin:1985cu} with  $(+----)$ signature. 
Motivated by the $D=3+2$ split of the solutions that we are interested in, a convenient basis for the Clifford algebra is given by 
\begin{align}\label{eq:5d_clifford_def}
\gamma^0 = & \sigma^2 \otimes \sigma^3, \qquad  \gamma^1 =  i\sigma^3 \otimes \sigma^3, \qquad
\gamma^2 = i\sigma^1 \otimes \sigma^3\,, \nn
\gamma^3 =&  {1} \otimes i  \sigma^1, \qquad
\gamma^4 =   {1} \otimes  i  \sigma^2, 
\end{align}
with $\gamma^{01234}=-1$. Defining $C =-C^{-1}=-i \gamma^0 \gamma^4$ and 
$B=+B^{-1}= -i\gamma^4$, we have $C \gamma_A C^{-1}=  + \gamma_A^T$ and
$B \gamma_A B^{-1}=  + \gamma_A^*$. 
As in \cite{Gunaydin:1985cu} a symplectic Majorana spinor $\epsilon^a$, with $a=1,2$ satisfies
$\epsilon^a=C(\bar \epsilon^a)^T$, with $\bar \epsilon^a=(\epsilon_a)^\dagger\gamma_0$, where 
we raise and lower symplectic indices using 
$\Omega^{ab} = \Omega^{[ab]} = (\Omega_{ba})^*$, with $\Omega_{ab} \Omega^{bc} = \delta_a^c$.
For the $SO(7)$ gamma matrices used in \cite{Gunaydin:1985cu}, we take the explicit realisation given in appendix C.1 of \cite{Freedman:1999gp}.

Our starting point is the consistent Kaluza-Klein truncation of maximal $SO(6)$ gauged supergravity discussed in \cite{Bobev:2010de},
extending \cite{Khavaev:2000gb}. It can be constructed in a two step procedure. One first considers
a $\mathbb{Z}_2\times\mathbb{Z}_2\subset SO(6)$ invariant sector which gives rises to an $\mathcal{N}=2$ gauged supergravity theory 
with two vector multiplets and 4 hypermultliplets (18 scalars in total). 
Then one utilises an additional $\mathbb{Z}_4\subset SO(6)\times SL(2)$, as in 
\cite{Khavaev:2000gb}, to further truncate the hypermultiplets. One is then left with a theory whose bosonic content consists
of a metric, three gauge fields $A^{(1)},A^{(2)},A^{(3)}$, two real and neutral scalars $\alpha,\beta$ that live in the $\mathcal{N}=2$ vector multiplets, and four complex and charged scalar fields $\zeta_j=e^{i\theta_j}\tanh\varphi_j$ that are maintained from the hypermultiplets and parametrise\footnote{This should not be confused with the model of \cite{Bobev:2016nua}, also used in \cite{Arav:2020obl}, with scalars parametrising the same coset.}
 the coset $[SU(1,1)/U(1)]^4$.
 
We have obtained this truncation using an ansatz for the $SO(6)$ gauged supergravity fields as in  
\cite{Khavaev:2000gb}, but with some adjustments (also slightly differing from \cite{Suh:2018vbp}).
The 42 scalars parametrise the non-compact coset space $E_{6(6)}/USp(8)$ and can be characterised by
a $27\times 27$ matrix $\mathcal{V}_{AB}{}^{ab}$. The action of the Lie algebra of $E_{6(6)}$ can be displayed using
a basis adapted to the maximal subgroup $SL(6,\mathbb{R})\times SL(2,\mathbb{R})\subset E_{6(6)}$. Specifically,
the infinitesimal $E_{6(6)}$ transformations acting on the vector space $\hat z^{AB}=(\hat z_{IJ},\hat z^{I\alpha})$, with $I,J,...=1,\dots,6$ and 
$\alpha,\beta,...=1,2$ can be written (consistent with (A.35)and (A.36) of \cite{Gunaydin:1985cu}) 
\begin{align}
\delta \hat z_{IJ}&=-\Lambda^K{}_I\hat z_{KJ}-\Lambda^K{}_J \hat z_{IK}+\sqrt{2}\Sigma_{IJK\beta}\hat z^{K\beta}\,,\nn
\delta \hat z^{I\alpha}&=\frac{1}{\sqrt{2}}\Sigma^{KLI\alpha}\hat z_{KL}+\Lambda^I{}_K\hat z^{K\alpha}+\Lambda^\alpha{}_\beta \hat z^{I\beta}\,.
\end{align}
Here, $\Lambda^I{}_J$, $\Lambda^\alpha{}_\beta$ are real, traceless and generate $SL(6,\mathbb{R})$ and
$SL(2,\mathbb{R})$, respectively, and $\Sigma_{IJK\alpha}$ is real and antisymmetric in $IJK$.

Similar to \cite{Khavaev:2000gb}, we next introduce 6 real Cartesian coordinates $x^I$ and 2 real Cartesian
coordinates $y^\alpha$. We define the differential form\footnote{Note that in \cite{Khavaev:2000gb} there was a factor of $1/12$ here instead; the difference is compensated in our normalisation in \eqref{sigmaupsilon}.}
\begin{align}
\Sigma=\frac{1}{6}\Sigma_{IJK\alpha}dx^I\wedge dx^J\wedge dx^K\wedge dy^\alpha\,.
\end{align}
Next we introduce four complex coordinates: $z_1 = x^1 + i x^2$, $z_2= x^3 - ix^4$, $z_3 = x^5 - ix^6$ and $z_4 = y^1 - i y^2$ (note the difference in $z_4$ from \cite{Khavaev:2000gb}). We then define the following 4-forms (note again the differences from \cite{Khavaev:2000gb}):
\begin{align}
\Upsilon_1 &\equiv - dz_1 \wedge dz_2 \wedge dz_3 \wedge dz_4\, , 
&
\Upsilon_2 &\equiv - dz_1 \wedge d\bar{z}_2 \wedge d\bar{z}_3 \wedge dz_4\, ,  \nn
\Upsilon_3 &\equiv - d\bar{z}_1 \wedge dz_2 \wedge d\bar{z}_3 \wedge dz_4\, , 
&
\Upsilon_4 &\equiv - d\bar{z}_1 \wedge d\bar{z}_2 \wedge dz_3 \wedge dz_4\, .
\end{align}
The parametrisation of the scalar coset in the truncated theory is then obtained by writing
the $\Sigma$ tensor as (note the normalization used here):
\begin{equation}\label{sigmaupsilon}
\Sigma = \frac{1}{4\sqrt{2}} \left( \sum_{i=1}^4 \zeta_i \Upsilon_i + \text{c.c.} \right)\, ,
\end{equation}
where $\zeta_i=\varphi_i e^{i \theta_i}$ are 4 complex scalars, and 
the $\Lambda$ tensor as
\begin{equation}
\Lambda = \operatorname{diag}\left( -\alpha + \beta, -\alpha + \beta, -\alpha - \beta, -\alpha - \beta, 2\alpha, 2\alpha  \right)\, ,
\end{equation}
where $\alpha,\beta$ are two real scalars. The $SO(6)$ gauge fields lie in a $U(1)^3\subset SO(6)$ sector
and specifically, we take
\begin{equation}
A =
\begin{pmatrix}
0 & -A^{(1)} & & & & & & \\
A^{(1)} & 0  & & & & & & \\
& & 0 & A^{(2)} & & & & \\
& & -A^{(2)} & 0 & & & & \\
& & & & & & 0 & -A^{(3)} \\
& & & & & & A^{(3)} & 0
\end{pmatrix} \, .
\end{equation}
Finally, we set the two-forms to zero.
Note these definitions are consistent with invariance under $ \mathbb{Z}_2 \times \mathbb{Z}_2 \times \mathbb{Z}_4 $ as discussed in \cite{Khavaev:2000gb}, which ensures the consistency of the truncation. The 27-bein and other supergravity tensors are then derived as explained in \cite{Gunaydin:1985cu}.

The bosonic part of the Lagrangian of the resulting truncated model, in a mostly minus signature, is given by 
\begin{align}\label{model1}
\mathcal{L} &= -\tfrac{1}{4} R
 +  \sum_{j=1}^4 \Big[\tfrac{1}{2}(\partial \varphi_j)^2+  \tfrac{1}{8}\sinh^2 2\varphi_j  \left(D\theta_j \right)^2\Big]
 + 3 (\partial \alpha)^2 + (\partial \beta)^2   - \mathcal{P} \nn
&
 - \tfrac{1}{4}\left[ e^{4\alpha-4\beta} F^{(1)}_{\mu\nu}F^{(1)\mu\nu}
+ e^{4\alpha+4\beta} F^{(2)}_{\mu\nu}F^{(2)\mu\nu}
+ e^{-8\alpha} F^{(3)}_{\mu\nu}F^{(3)\mu\nu} \right]
+\tfrac{1}{2}\epsilon^{\mu\nu\rho\sigma\delta}F^{(1)}_{\mu\nu}F^{(2)}_{\rho\sigma}A^{(3)}_\delta\,,
\end{align} 
where 
\begin{align}\label{dthetas}
 D\theta_1&=\left(d \theta_1 +g A^{(1)} + gA^{(2)} - gA^{(3)} \right)\,,\nn
  D\theta_2&=\left(d \theta_2 + gA^{(1)} - gA^{(2)}_\mu + gA^{(3)}  \right)\,,\nn
 D\theta_3&=\left(d \theta_3 - gA^{(1)} + gA^{(2)} + gA^{(3)} \right)\,,\nn
 D\theta_4&=\left(d \theta_4 -gA^{(1)}- gA^{(2)} - gA^{(3)} \right)\,.
\end{align}
The scalar potential $\mathcal{P}$ is given by
\begin{equation}\label{model2}
\mathcal{P} = \frac{g^2}{8} \left[\sum_{j=1}^4 \left( \frac{\partial W}{\partial \varphi_j} \right)^2 + \frac{1}{6}  \left( \frac{\partial W}{\partial \alpha} \right)^2 + \frac{1}{2} \left( \frac{\partial W}{\partial \beta} \right)^2 \right] - \frac{g^2}{3} W^2 \, ,
\end{equation}
where $W$ is the superpotential defined by
\begin{align}\label{superpot}
W = -\frac{1}{4} \Big[ 
 &( e^{-2\alpha-2\beta}+e^{-2\alpha+2\beta } - e^{ 4\alpha } ) \cosh2\varphi_1  
 + ( -e^{-2\alpha-2\beta}+e^{-2\alpha+2\beta } + e^{ 4\alpha } ) \cosh2\varphi_2  \nn
 +& 
 ( e^{-2\alpha-2\beta}-e^{-2\alpha+2\beta } + e^{ 4\alpha } ) \cosh2\varphi_3  
 + ( e^{-2\alpha-2\beta}+e^{-2\alpha+2\beta } + e^{ 4\alpha } ) \cosh2\varphi_4  
  \Big] \, .
\end{align}
The model has various discrete symmetries, generalising those mentioned in section 3 of \cite{Arav:2020obl}, for example.
These include
\begin{align}
\beta\to-\beta, \quad \zeta_2\leftrightarrow \zeta_3,\quad A^{(1)}\leftrightarrow  A^{(2)}\,,
\end{align}
and
\begin{align}
\alpha\to\frac{1}{2}(-\alpha+\beta),\quad
\beta\to \frac{1}{2}(\beta+3\alpha), \quad \zeta_1\leftrightarrow \zeta_3,\quad A^{(1)}\leftrightarrow  A^{(3)}\,.
\end{align}

The model admits the maximally supersymmetric $AdS_5$ vacuum solution with the $AdS_5$ metric having radius squared equal to $4/g^2$.
Within the dual $\mathcal{N}=4$ SYM theory we identify the scalar fields $\alpha,\beta$ with bosonic mass deformations living in ${\bf 20}'$ of $SO(6)$ and $\zeta_j$ with fermionic mass deformations living in the ${\bf 10}$ of $SO(6)$.
If $X_a$ are the six real scalars and $\lambda_j$ the four fermions of $\mathcal{N}=4$ SYM theory then, schematically, 
we have \cite{Khavaev:2000gb}: 
\begin{align}\label{opfieldmapz}
\Delta=2:\qquad \qquad  \alpha &\quad \leftrightarrow \quad \tr(X_1^2+X_2^2+X_3^2+X_4^2-X_5^2-X_6^2)\,,\nn
   \beta &\quad \leftrightarrow \quad \tr(X_1^2+X_2^2-X_3^2-X_4^2)\,,\nn
\Delta=3:\qquad \qquad  \zeta_j &\quad \leftrightarrow \quad \tr(\lambda_j\lambda_j+\text{cubic in $X_a$})\,, \qquad j=1, 2, 3,4 \,, 
\end{align}
where $\Delta$ is the conformal scaling dimension of the operator.

The truncated model \eqref{model1}-\eqref{superpot} is not supersymmetric as it has incomplete $\mathcal{N}=2$ hypermultiplets. However, we can determine the conditions that need to be satisfied in order that a solution preserves some of this
$\mathcal{N}=2$ supersymmetry. We start with the $\mathcal{N}=8$ supersymmetry variations of the maximal
theory \cite{Gunaydin:1985cu} given by 
\begin{align}\label{gaugvar0}
\delta\psi_{\mu a} &=\nabla_\mu \epsilon_a +Q_{\mu a}{}^b\epsilon_b- \frac{1}{6}\, g\, W_{ab} \gamma_\mu \epsilon^b  - \frac{1}{6}\, H_{\nu \rho\, ab} \big(\gamma^{\nu \rho} \gamma_\mu + 2 \gamma^\nu \delta^\rho_\mu  \big)\, \epsilon^b   \,, \nn
\frac{1}{\sqrt 2}\delta\chi_{abc} &= \gamma^\mu P_{\mu\, abcd} \, \epsilon^d - \frac{1}{2} \, g\, A_{dabc}\, \epsilon^d  - \frac{3}{4}\, \gamma^{\mu \nu}  H_{\mu \nu\, [ab}  \, \epsilon_{c]}  \,,  
\end{align} 
with $\nabla_\mu=\partial_\mu+\tfrac{1}{4}\omega_{\mu ab}\gamma^{ab}$. We have
$W_{ab}=W_{ba}$, $H_{\nu\rho ab}=H_{\nu\rho [ab]}$, $A_{abcd}=A_{a[bcd]}$ and 
$Q_{\mu ab}=Q_{\mu (ab)}$, where $Q_{\mu ab}\equiv \Omega _{bc}Q_{\mu a}{}^c$ and 
$\Omega_{ab}$ is the symplectic form, which we choose to be as in (C.5) of \cite{Freedman:1999gp}. 
As in \cite{Freedman:1999gp,Bobev:2010de}, the $\mathcal{N}=2$ supersymmetry variations can be obtained by considering
the eigenvalues of the $W_{ab}$ tensor.  Indeed there is a symplectic pair of such spinors $\eta^a_{(k)}$, $k=1,2$, satisfying
\begin{align}
W_{ab} \, \eta^b_{(k)} &~=~W\,  \eta^a_{(k)} \,, \qquad k = 1,2 \,,
\label{evecs}
\end{align} 
with $W$ as in \eqref{superpot}. Explicitly we have 
\begin{align}
\eta^b_{(1)}&= \frac{1}{2} (-1, 0, 1, 0, 0, 1, 0, 1) \,,\nn
\eta^b_{(2)}&= \frac{1}{2} (0, -1, 0, -1, -1, 0, 1, 0) \,.
\end{align}
Note that we
have $\Omega_{ab} \,  \eta^b_{(1)}=-\eta^a_{(2)} $, $\Omega_{ab} \,  \eta^b_{(2)} = \eta^a_{(1)}$ and 
$\eta^a_{(k)}\eta^a_{(l)}=\delta_{(k)(l)}$. We can also define
\begin{align}
Q_\mu&\equiv Q_{\mu a}{}^b\eta^a_{(1)}\eta^b_{(2)}= -Q_{\mu a}{}^b\eta^a_{(2)}\eta^b_{(1)}\,,\nn
 H_{\nu\rho }&\equiv H_{\nu\rho ab}\eta^a_{(1)}\eta^b_{(2)}\,,
\end{align}
and we note $ Q_{\mu a}{}^b\eta^a_{(1)}\eta^b_{(1)}=Q_{\mu a}{}^b\eta^a_{(2)}\eta^b_{(2)}=0$. The 
$\mathcal{N}=2$ supersymmetry parameters are then given by a pair of symplectic Majorana spinors $\hat \epsilon_1$, 
$\hat \epsilon_2$, defined by
\begin{align}
\epsilon^a=\eta^a_{(1)}\hat\epsilon_1+\eta^a_{(2)}\hat\epsilon_2,\qquad\Rightarrow\qquad
\epsilon_a\equiv \Omega_{ab}\epsilon^b=-\eta^a_{(2)}\hat\epsilon_1+\eta^a_{(1)}\hat\epsilon_2.
 \end{align}
Finally, it is convenient to parametrise the $\mathcal{N}=2$ supersymmetry variations by a complex Dirac spinor defined by
$\epsilon\equiv \hat \epsilon_1+i\hat \epsilon_2$.

Using these ingredients we find that the vanishing of the gravitino variations in \eqref{gaugvar0} lead to the $\mathcal{N}=2$
supersymmetry conditions
\begin{align}\label{firstgravpass2}
&\Big(\nabla_\mu -iQ_\mu-\frac{ig}{6}W\gamma_\mu
-\frac{1}{12}H_{\nu\rho}(\gamma^{\nu\rho}\gamma_\mu+2\gamma^\nu\delta^\rho_\mu)\Big)\epsilon=0\,,
\end{align}
with $W$ as in \eqref{superpot},
\begin{align}\label{HQdefs}
H_{\mu\nu} =& e^{2\alpha-2\beta} F^{(1)}_{\mu\nu} + e^{2\alpha + 2\beta} F^{(2)}_{\mu\nu} + e^{-4\alpha} F^{(3)}_{\mu\nu}  \, ,\nn
Q_\mu=&-\frac{g}{2}(A^{(1)}_\mu+A^{(2)}_\mu+A^{(3)}_\mu)
-\frac{1}{4}(\cosh2\varphi_1-1)D_\mu\theta_1
-\frac{1}{4}(\cosh2\varphi_2-1)D_\mu\theta_2
\nn
&
-\frac{1}{4}(\cosh2\varphi_3-1)D_\mu\theta_3
+\frac{1}{4}(\cosh2\varphi_4-1)D_\mu\theta_4\,,
\end{align}
and note the sign of the last term.

Carrying out a similar procedure for the $\mathcal{N}=8$ gaugino variations in \eqref{gaugvar0}, we find the supersymmetry conditions
associated with $\mathcal{N}=2$ gaugino variations are given by
\begin{align}\label{gauginovars}
[\gamma^\mu\partial_\mu \alpha +\frac{ig}{12}\partial_\alpha W-\frac{1}{12}(e^{2\alpha-2\beta} F^{(1)}_{\mu\nu} + e^{2\alpha+2\beta} F^{(2)}_{\mu\nu} - 2 e^{-4\alpha} F^{(3)}_{\mu\nu} )\gamma^{\mu\nu}]\epsilon&=0\,,\nn
{}[\gamma^\mu\partial_\mu \beta +\frac{ig}{4}\partial_\beta W-\frac{1}{4}(-e^{2\alpha-2\beta} F^{(1)}_{\mu\nu} +e^{2\alpha+2\beta} F^{(2)}_{\mu\nu})\gamma^{\mu\nu}]\epsilon&=0\,,
\end{align}
while those for the $\mathcal{N}=2$ hyperino variations are given by
\begin{align}\label{varhyperino}
{}[\gamma^\mu\partial_\mu \varphi_j +\frac{ig}{2}\partial_{\varphi_j}W+i\partial_{\varphi_j} Q_\mu\gamma^\mu]\epsilon=0\,.
\end{align}

If we set the four charged scalars to zero, $\zeta_j=0$, we obtain the STU model; see section \ref{stutrunc} below.
From the STU model, we obtain the truncation to minimal $D=5$ gauged supergravity by setting all gauge fields equal
$A^{(1)}=A^{(2)}=A^{(3)}$.
There is a second way to reduce to minimal gauged supergravity associated with the LS $AdS_5$ solution as we discuss in section
\ref{mingstrunc}.

We also note that there is an overlap with the ten scalar model of \cite{Bobev:2016nua}. If we set
$\zeta_j$ to be purely imaginary we obtain a six scalar model. This can be obtained from \cite{Bobev:2016nua}
by setting their $\alpha_i=0$ (with the $\alpha_i$ associated with boson masses in $\mathcal{N}=4$ SYM of the form
$\tr(X_1^2-X_2^2)$, $\tr(X_3^2-X_4^2)$ and $\tr(X_5^2-X_6^2)$) as well
as $\varphi=0$ (dual to the gauge coupling). Note that one should identify $\alpha,\beta$ here with $-\beta_1,\pm\beta_2$ in
\cite{Bobev:2016nua}.

\subsection{The Extended LS Subtruncation}

This sub-truncation, which was also used in \cite{Bobev:2014jva}, is the one that is utilised in the paper. It is obtained by further setting three of the charged fields to zero:
\begin{equation}\label{extlstrunc}
\varphi_2 = \varphi_3 = \varphi_4 = \theta_2 = \theta_3 = \theta_4 = 0 \, .
\end{equation}
It keeps three gauge fields, one complex scalar\footnote{In the text we have dropped the subscript on $\varphi_1$ and $\theta_1$.}, $\zeta_1=\varphi_1e^{i\theta_1}$ along with the two real scalars $\alpha,\beta$.
This truncation is invariant under a $U(1) \times U(1)_R $ subgroup of the global $SU(4)$ symmetry of the maximal gauge supergravity.
More precisely, we first decompose $SU(3)\times U(1)_1\subset SU(4)$ and then further decompose $SU(2)\times U(1)_2\subset SU(3)$ and
define $U(1)_R$ to be the diagonal subgroup of $U(1)_1\times U(1)_2$. 

For this truncation we have
\begin{align}\label{WQLS}
W &= \frac{1}{4} \left[ -2 ( e^{-2\alpha-2\beta } + e^{ -2\alpha + 2\beta } ) \cosh^2\varphi_1 + e^{4\alpha} ( -3 + \cosh2\varphi_1 )  \right] \, ,\nn
Q_\mu&=-\frac{g}{2}(A^{(1)}_\mu+A^{(2)}_\mu+A^{(3)}_\mu)
-\frac{1}{4}(\cosh2\varphi_1-1)D_\mu\theta_1\,,
\end{align}
with $D\theta_1=\left(d \theta_1 +g A^{(1)} + gA^{(2)} - gA^{(3)} \right)$. Clearly $\zeta_1$ is uncharged under the linear combination of gauge transformations
$\delta (A^{(1)} ,A^{(2)}, A^{(3)})=\Lambda(1,-1,0)$ as well as
$\delta (A^{(1)} ,A^{(2)}, A^{(3)})=\Lambda(1,1,2)$. Moreover, from \eqref{rgfielddeftext} we
see that the former is a $U(1)$ flavour symmetry, while the latter is a $U(1)$ $R$-symmetry.

This truncation contains the supersymmetric LS $AdS_5$ fixed point solutions given by 
\begin{align}
e^{6\alpha}=2,\qquad \cosh(2\varphi_1)=\frac{5}{3}\,,
\end{align}
with $\beta=0$ and vanishing gauge fields.
These solutions preserves $SU(2)\times U(1)_R$ symmetry.
If we further truncate by setting $A^{(1)}=A^{(2)}$ as well as $\beta=0$ then we obtain a model that preserves $SU(2)\times U(1)_R$ symmetry  which
contains the metric, two gauge fields, a complex scalar $\zeta_1$ and a real scalar field $\alpha$, and also contains the LS fixed points.
Further setting the gauge fields to zero leads to an $SU(2)\times U(1)_R$ invariant model with metric, and two real scalar fields, $\alpha,\varphi_1$, which was used in \cite{Freedman:1999gp} to construct the RG flow from the maximally supersymmetric vacuum to the LS fixed point. One might refer to this as the LS sub-truncation.
The model of interest here, defined by \eqref{extlstrunc}, preserves a smaller amount of the global symmetry, namely $U(1) \times U(1)_R $, and more fields, so we refer to it as the \emph{extended LS sub-truncation}.

From the extended LS truncation we can further reduce in two different ways to minimal $D=5$ gauged supergravity as we discuss
in sections \ref{stutrunc} and \ref{mingstrunc}.

There is also overlap with the model considered in section 3.1 of \cite{Arav:2020obl}.

\subsection{The $\mathcal{N}=2^*$ Subtruncation}

This sub-truncation is obtained by setting
\begin{align}\label{neqtwoconds}
\varphi_1&=\varphi_4=\theta_1=\theta_4=0 \, , \quad \beta = 0 
\, ,  \nn
\varphi_2&=\varphi_3\,, \quad 
\theta_2=\theta_3\,,
\quad
A^{(1)}_\mu = A^{(2)}_\mu \, . \quad
\end{align}
It thus keeps two gauge fields, one complex scalar $\zeta_2$ and one real scalar $\alpha$.
For this model we have
\begin{align}
W &= - \frac{1}{2} ( 2 e^{-2\alpha} + e^{4\alpha} \cosh 2\varphi_2 ) \, ,\nn
Q_\mu&=-\frac{g}{2}(2A^{(1)}_\mu+A^{(3)}_\mu)
-\frac{1}{2}(\cosh2\varphi_2-1)D_\mu\theta_2\,,
\end{align}
with $D\theta_2=\left(d \theta_2 +gA^{(3)}  \right)$.

When we set $A^{(1)}=0$, the truncation is invariant under an $SU(2)_R \times U(1)'$ subgroup of the $SU(4)$ symmetry of the maximal gauged supergravity.
More precisely, we first decompose $SU(2)_1\times SU(2)_2\times U(1)\subset SU(4)$, with $SU(2)_1$ and $SU(2)_2$
rotating $\zeta_{1,4}$ and $\zeta_{2,3}$, respectively, and then the $SU(2)_R$ factor is the $SU(2)_1$ factor. The $U(1)'$ factor is a subgroup of $SU(2)_2$ and the gauge field $A^{(3)}$ is associated with the $U(1)$ factor.
With $A^{(1)}\ne 0$ the truncation is invariant under an $U(1)_R \times U(1)'$ 

For this particular truncation any solution  to the equations of motion with $A^{(1)}=0$ that satisfies the supersymmetry equations
\eqref{firstgravpass2}-\eqref{varhyperino} will preserve twice as much supersymmetry in the full $\mathcal{N}=8$ gauged supergravity theory. This is indicated by calculating the $W$ tensors of \cite{Gunaydin:1985cu}.

The truncated theory \eqref{neqtwoconds} overlaps with the $SU(2)\times U(1)$ invariant sub-truncation of maximal gauged supergravity \cite{Pilch:2000fu}, which 
is an $\mathcal{N}=2$ supergravity theory coupled to one vector multiplet and one hypermultiplet, and there
is also overlap with the model considered in section 3.3 of \cite{Arav:2020obl}.

\subsection{The $SO(3)$ Subtruncation}

This sub-truncation is obtained by taking:
\begin{align}
\varphi_1&=\varphi_2=\varphi_3 \, , \quad
\theta_1=\theta_2=\theta_3 \, , \nn
A^{(1)}_\mu &= A^{(2)}_\mu = A^{(3)}_\mu \, , \quad
\alpha=\beta = 0 \, . 
\end{align}
This keeps one gauge field $A^{(1)}_\mu$ and two complex scalar fields $\zeta_1,\zeta_4$.
The truncation is invariant under an $SO(3)\subset SU(3)\subset SU(4)$ subgroup.
For this model we have
\begin{align}
W &= - \frac{3}{4} ( \cosh2\varphi_1 + \cosh2\varphi_4 ) \, .\nn
Q_\mu&=-\frac{3g}{2}A^{(1)}_\mu
-\frac{3}{4}(\cosh2\varphi_1-1)D_\mu\theta_1+\frac{1}{4}(\cosh2\varphi_4-1)D_\mu\theta_4\,,
\end{align}
with $D\theta_1=\left(d \theta_1 +gA^{(1)}  \right)$
and $D\theta_4=\left(d \theta_4 -3gA^{(1)}  \right)$.

There is overlap with this model and the model considered in section 3.2 of \cite{Arav:2020obl}.

\subsection{Truncation to the STU model}\label{stutrunc}
The truncation to the STU model to minimal gauged supergravity is obtained by setting all of the charged scalars to zero:
\begin{align}
\varphi_i = \theta_i  = 0\,,
\end{align}
to get a $D=5$ Lagrangian with $\mathcal{P}$  given by
\begin{equation}\label{model2stu}
\mathcal{P} = -4g^2(e^{-4\alpha}+e^{2\alpha-2\beta}+e^{2\alpha+2\beta})\,.
\end{equation}
If we set $g=2$ and take $F^{(i)}\to \frac{1}{2}F^{(I)}$ we get the same normalisations of the STU model used in \cite{Ferrero:2021etw}, provided that we identify the neutral scalars $\alpha,\beta$ here with, respectively, $\varphi_1/(2\sqrt{6}),-\varphi_2/(2\sqrt{2})$ there.

Further setting 
\begin{align}
\alpha=\beta=0,\quad
 A^{(1)}&=A^{(2)}=A^{(3)}\,,
\end{align}
we obtain a $D=5$ Lagrangian given by
\begin{align}\label{modelstumin}
\mathcal{L} =
\tfrac{1}{4}\left[-R
 +{3g^2} - {3}F^{(3)}_{\mu\nu}F^{(3)\mu\nu}
+2\epsilon^{\mu\nu\rho\sigma\delta}F^{(3)}_{\mu\nu}F^{(3)}_{\rho\sigma}A^{(3)}_\delta\right]\,.
\end{align} 
This can also be obtained from the extended LS truncation.
If we set $g=2$ and take $F^{(3)}\to \frac{1}{3}F$ we get the same normalisations of the minimal $D=5$ gauged supergravity
model used in \cite{Ferrero:2020laf}.

\subsection{Truncation to minimal gauged supergravity via LS }\label{mingstrunc}
An alternative way to truncate to minimal $D=5$ gauged supergravity is associated with the LS $AdS_5$ solution. We set
\begin{align}
e^{6\alpha}&=2,\quad  \cosh 2\varphi_1=\frac{5}{3},\quad 
\beta=0, \quad \varphi_2 = \varphi_3 = \varphi_4 = \theta_i  = 0\,,\nn
 A^{(1)}&=A^{(2)}=\frac{1}{2}A^{(3)}\,,
\end{align}
to obtain
\begin{align}\label{lsmin}
\mathcal{L} &=\tfrac{1}{4} \left[ -R +\tfrac{2^{10/3}g^2}{3}  - \tfrac{3}{ 2^{4/3}} F^{(3)}_{\mu\nu}F^{(3)\mu\nu} 
+\tfrac{1}{2}\epsilon^{\mu\nu\rho\sigma\delta}F^{(3)}_{\mu\nu}F^{(3)}_{\rho\sigma}A^{(3)}_\delta\right]\,.
\end{align} 
This model can also be obtained from the extended LS truncation.
If we set $g=\frac{3}{2^{2/3}}$ and take $F^{(3)}\to \frac{2^{2/3}}{3}F$ we get the same normalisations of minimal $D=5$ gauged 
supergravity
as used in \cite{Ferrero:2020laf}.

\section{Supersymmetry variations}\label{susyvars}
\subsection{Derivation of the BPS equations}\label{sec:b1}
Here we analyse the supersymmetry variations \eqref{firstgravpass2} for the $AdS_3$ ansatz given in 
\eqref{ads3ans}. The analysis is somewhat similar to the derivation of BPS equations for backgrounds with the $AdS_3$ factor
replaced with an $S^3$ that were studied in \cite{Bobev:2010de}.

We use the orthonormal frame given in \eqref{orthframe}.
We also use the gamma matrices given in \eqref{eq:5d_clifford_def} and
write the Killing spinor as
\begin{align}
\epsilon=\psi\otimes\chi\,,
\end{align}
with $\psi$ a two component spinor on $AdS_3$ which satisfies
\begin{align}
D_m\psi=\frac{i}{2}\kappa\Gamma_m\psi\,,
\end{align}
where $\kappa=\pm 1$ and
 $\Gamma^m=(\sigma^2,  i\sigma^3, i\sigma^1)$ are gamma matrices in $D=3$, with mostly minus signature.

By considering the components of the gravitino variation \eqref{firstgravpass2} that are tangent to the $AdS_3$ directions we deduce
\begin{align}
\left[ -\left(3 \kappa e^{-V}+ {H}_{34}  \right) \gamma^{34} + {3 V'}{f^{-1}}  \gamma^3 \right] \epsilon = i g W\epsilon\,.
\end{align}
For this to have nontrivial solutions the left hand side must have eigenvalue $+igW$, which requires that the two coefficients live on a circle and so we can write
\begin{align}\label{projeps}
\left[ \cos\xi \gamma^{34} + \sin\xi \gamma^3 \right] \epsilon = +i \epsilon\,,
\end{align}
where
\begin{align}\label{angdefs}
-3 \kappa e^{-V} - H_{34}= gW \cos\xi\,, \qquad  3 V'f^{-1} = g W \sin\xi\,.
\end{align}
The projection condition can then be solved by writing
\begin{align}
\epsilon = e^{\frac{\xi}{2} \gamma^4} \eta,\qquad \gamma^3 \eta = +i \gamma^4 \eta\,.
\end{align}
Notice from \eqref{angdefs} we have $\partial_z\xi=0$. 
We also observe that at $\xi=0,\pi$ the spinors have a definite chirality with
respect to $\gamma^{34}=-i(1\otimes \sigma_3)$:
\begin{align}\label{kschirality}
\xi=0,\pi\qquad \gamma^{34}\epsilon=\pm i\epsilon.
\end{align}

We next consider the components of the gravitino variation \eqref{firstgravpass2} in the $y$ direction. After a little work we can write this as
\begin{align}\label{ycond}
\Big[\partial_y-\frac{1}{2}V'+\frac{1}{2}\left(  \partial_y\xi+f H_{34}+\kappa f e^{-V} \right)\gamma^4\Big]\eta=0\,,
\end{align}
where we used \eqref{angdefs}. From the components in the $z$ direction we deduce 
\begin{align}\label{zcond}
\Big[\partial_z-iQ_z&+\frac{i}{2}f^{-1}h'\cos\xi-\frac{i}{3}H_{34}h \sin\xi  \nn
&+i\left( -\frac{1}{2}f^{-1}h'\sin\xi+\frac{gWh}{6}-\frac{1}{3}H_{34}h\cos\xi        \right)\gamma^4\Big]\eta=0\,.
\end{align}

An expression of the form $(a_1+a_2\gamma^4)\eta=0$ implies that $a_1^2+a_2^2=0$. Thus,
from \eqref{ycond}, \eqref{zcond} we can deduce
\begin{align}
\eta=e^{V/2}e^{is z}\eta_0\,,
\end{align}
where $\eta_0$ is independent of $y$ and $z$, along with the conditions
\begin{align}\label{gravitinoeqs}
\partial_y\xi+f H_{34}+\kappa f e^{-V}&=0\,,\nn
(s-Q_z)+\frac{1}{2}f^{-1}h'\cos\xi-\frac{1}{3}H_{34}h \sin\xi &=0\,,\nn
-\frac{1}{2}f^{-1}h'\sin\xi+\frac{gWh}{6}-\frac{1}{3}H_{34}h\cos\xi   &=0\,.
\end{align}
From these we deduce
\begin{align}\label{hprimeH}
f^{-1}h'&=\frac{gWh}{3}\sin\xi-2(s-Q_z)\cos\xi\,,\nn
hH_{34}&=\frac{gWh}{2}\cos\xi+3(s-Q_z)\sin\xi\,,
\end{align}
and from the first of \eqref{angdefs} we then have
\begin{align}\label{kappaagain}
(s-Q_z)\sin\xi=-\frac{1}{2}g  Wh\cos\xi-\kappa h e^{-V}\,,
\end{align}
and hence
\begin{align}
H_{34}&=-gW\cos\xi-3\kappa e^{-V}\,,\nn
f^{-1}\partial_y\xi&=gW\cos\xi+2\kappa e^{-V}\,.
\end{align}
When $\sin\xi\ne 0$, we can solve for $(s-Q_z)$ and also write 
\begin{align}
f^{-1}\frac{h'}{h}\sin\xi&=2\kappa e^{-V}\cos\xi+\frac{gW}{3}(1+2\cos^2\xi)\,.
\end{align}

Proceeding in the same way, from the gaugino variations we deduce
\begin{align}\label{gaugino1}
f^{-1}\alpha'+\frac{g}{12}\partial_\alpha W\sin\xi&=0\,,\nn
f^{-1}\beta'+\frac{g}{4}\partial_\beta W\sin\xi&=0\,,
\end{align}
and
\begin{align}
g\partial_\alpha W\cos\xi-2(e^{2\alpha-2\beta} F^{(1)}_{34} + e^{2\alpha+2\beta} F^{(2)}_{34} - 2 e^{-4\alpha} F^{(3)}_{34} )&=0\,,\nn
g\partial_\beta W\cos\xi-2(-e^{2\alpha-2\beta} F^{(1)}_{34} +e^{2\alpha+2\beta} F^{(2)}_{34})&=0\,.
\end{align}
From these last two expressions, recalling the definition of $H_{34}$ from \eqref{HQdefs}, and using \eqref{hprimeH} as well
as the expression for $(s-Q_z)$ in \eqref{kappaagain}, we obtain the following expressions for the components of the field strengths in the orthonormal frame 
\begin{align}
e^{2\alpha-2\beta}F^{(1)}_{34}&=\frac{g}{6}[W+\frac{1}{2}(\partial_\alpha W-3\partial_\beta W)]\cos\xi+h^{-1}(s-Q_z)\sin\xi\,,\nn
&=-\frac{g}{12}[4W-\partial_\alpha W+3\partial_\beta W]\cos\xi-\kappa e^{-V}\,,\nn
e^{2\alpha+2\beta}F^{(2)}_{34}
&=-\frac{g}{12}[4W-\partial_\alpha W-3\partial_\beta W]\cos\xi-\kappa e^{-V}\,,\nn
e^{-4\alpha}F^{(3)}_{34}
&=-\frac{g}{6}[2W+\partial_\alpha W]\cos\xi-\kappa e^{-V}\,.
\end{align}

From the hyperino equations \eqref{varhyperino} we obtain
\begin{align}\label{hyperino1}
f^{-1}\partial_y\varphi_j+\frac{g}{2}\partial_{\varphi_j}W\sin\xi+\partial_{\varphi_j} Q_z\cos\xi  h^{-1}=0\,,\nn
\frac{g}{2}\partial_{\varphi_j}W\cos\xi-\partial_{\varphi_j} Q_z \sin\xi h^{-1}=0\,.
\end{align}
Notice that for each $\varphi_j$ that is identically zero, $\varphi_j\equiv 0$ (e.g. three of the four scalars in the extended LS truncation) these equations are trivially satisfied and impose no constraints.

{\bf Summary:} When $\sin\xi\ne 0$, the BPS equations are given by the following first order equations
\begin{align}\label{summbbpsapp}
f^{-1}\xi'&=gW\cos\xi+2\kappa e^{-V}\,,\nn
f^{-1} V' &= \frac{g}{3} W \sin\xi,\nn
f^{-1}\alpha'&=-\frac{g}{12}\partial_\alpha W\sin\xi\,,\nn
f^{-1}\beta'&=-\frac{g}{4}\partial_\beta W\sin\xi\,,\nn
f^{-1}\varphi_j'&=-\frac{g}{2}\frac{\partial_{\varphi_j}W}{\sin\xi}\,,\nn
f^{-1}\frac{h'}{h}\sin\xi&=2\kappa e^{-V}\cos\xi+\frac{gW}{3}(1+2\cos^2\xi)\,,
\end{align}
along with two constraint equations
\begin{align}
(s-Q_z)\sin\xi&=-\frac{1}{2}g  Wh\cos\xi-\kappa h e^{-V}\,,\nn
\frac{g}{2}\partial_{\varphi_j}W\cos\xi&=\partial_{\varphi_j} Q_z \sin\xi h^{-1}\,.
\end{align}
The field strengths are given by
\begin{align}\label{fstrengthsbps}
e^{2\alpha-2\beta}F^{(1)}_{34}&=-\frac{g}{12}[4W-\partial_\alpha W+3\partial_\beta W]\cos\xi-\kappa e^{-V}\,,\nn
e^{2\alpha+2\beta}F^{(2)}_{34}
&=-\frac{g}{12}[4W-\partial_\alpha W-3\partial_\beta W]\cos\xi-\kappa e^{-V}\,,\nn
e^{-4\alpha}F^{(3)}_{34}
&=-\frac{g}{6}[2W+\partial_\alpha W]\cos\xi-\kappa e^{-V}\,,\nn
H_{34}&=-gW\cos\xi-3\kappa e^{-V}\,.
\end{align}

Along the BPS flow we have
\begin{align}
\partial_y W=-gf\sin\xi\Big[\frac{1}{12}(\partial_\alpha W)^2+\frac{1}{4}(\partial_\beta W)^2+\frac{1}{2\sin^2\xi}\sum_i(\partial_{\varphi^i} W)^2\Big]\,,
\end{align}
and hence we see that provided the sign of $f\sin\xi$ doesn't change, then $W$ is monotonic along the BPS flow.

Interestingly, by examining the derivative of $he^{-V}$ we can find an integral of the BPS equations:
\begin{align}\label{hemvszeq}
he^{-V}=k \sin\xi\,,
\end{align}
where $k$ is a constant. Eliminating the BPS equation for $h$ we can then write the remaining BPS equations in the form
\begin{align}\label{scderivsgen}
f^{-1}\xi'&=-2k^{-1}(s-Q_z)e^{-V}\,,\nn
f^{-1} V' &= \frac{g}{3} W \sin\xi\,,\nn
f^{-1}\alpha'&=-\frac{g}{12}\partial_\alpha W\sin\xi\,,\nn
f^{-1}\beta'&=-\frac{g}{4}\partial_\beta W\sin\xi\,,\nn
f^{-1}\varphi_j'&=-\frac{g}{2}\frac{\partial_{\varphi_j}W}{\sin\xi}\,,
\end{align}
and the two constraints can be written
\begin{align}\label{const2}
(s-Q_z)&=-k\big(\frac{1}{2}g  We^V\cos\xi+\kappa\big)\,, \nn
\frac{g}{2}\partial_{\varphi_j}W\cos\xi&=k^{-1}e^{-V} \partial_{\varphi_j} Q_z\,.
\end{align}

We now point out two additional observations concerning the BPS equations that are useful in the analysis of the main text. The first concerns the fluxes and the second the discrete symmetries of the BPS equations.
From the definition of $Q_\mu$ in \eqref{HQdefs} we can write
\begin{align}
\partial_{\varphi_j} Q_z=\mp \frac{1}{2}\sinh 2\varphi_j D_z\theta_j\,,
\end{align}
where the upper sign is for $j=1,2,3$ and the low sign for $j=4$. The second constraint equation
in \eqref{const2} is trivial for each specific $j$ for which $\varphi_j=0$. On the other hand
when all $\varphi_j \neq 0$ we can write the constraint in the form
\begin{equation}\label{consphizsero}
D_z \theta_j = \mp\frac{gk e^V \partial_{\varphi_j}W \cos\xi}{\sinh 2\varphi_j} \, ,
\end{equation}
and moreover the right hand side is independent of $\varphi_j$. This constraint must be consistent with the BPS equations, and therefore if one differentiates \eqref{consphizsero} one obtains an equality between a linear combination of the fluxes and a derivative of expressions involving the other fields.  Solving these we deduce that we can write
\begin{equation}\label{ifluxdefs}
F^{(i)}_{yz} = (a^{(i)})' = (\mathcal{I}^{(i)})' \, ,
\end{equation}
where
\begin{align}\label{eq:IntegratedFluxesExpr1app}
& \mathcal{I}^{(1)} \equiv \frac{1}{2} k e^V \cos\xi \, e^{-2\alpha+2\beta} \, , \nn
& \mathcal{I}^{(2)} \equiv \frac{1}{2} k e^V \cos\xi \, e^{-2\alpha-2\beta} \, , \nn
& \mathcal{I}^{(3)} \equiv \frac{1}{2} k e^V \cos\xi \, e^{4\alpha} \, .
\end{align}
Now \eqref{consphizsero} is not valid when some of the $\varphi_j=0$. However, we find
that the expressions for the fluxes in \eqref{ifluxdefs}, \eqref{eq:IntegratedFluxesExpr1app} still hold.

It is also helpful to observe that there are several symmetries of the BPS equations. The first is
\begin{align}\label{sym1}
h\to-h\,,\quad z\to -z\,,
\end{align}
with $Q_z\to -Q_z$, $s\to- s$, $a^{(i)}\to-a^{(i)}$, $k\to -k$ and $F^{(i)}_{34}\to +F^{(i)}_{34}$. 
This transformation leaves the frame invariant. We use this symmetry in the text to fix $h\ge 0$.
The second symmetry is 
\begin{align}\label{sym2}
\xi\to-\xi+\pi,\quad\kappa\to-\kappa,\quad z\to -z\,,
\end{align}
with $Q_z\to -Q_z$, $s\to- s$, $a^{(i)}\to-a^{(i)}$ and $F^{(i)}_{34}\to -F^{(i)}_{34}$. 
Also $\cos\xi\to-\cos\xi$, $\sin\xi\to +\sin\xi$.
This changes the frame $e^3\to -e^3$. Notice that this changes the sign of $\kappa$ and hence the chirality of the preserved supersymmetry of the $d=2$ SCFT i.e. whether it is $\mathcal{N}=(0,2)$ or
$\mathcal{N}=(2,0)$. We will not utilise this symmetry in our analysis in the text.
The third symmetry is
\begin{align}\label{sym3}
\xi\to -\xi,\quad y\to -y,\quad z\to -z
\end{align}
with $Q_z\to -Q_z$, $s\to- s$, $a^{(i)}\to-a^{(i)}$, $k\to -k$ and $F^{(i)}_{34}\to +F^{(i)}_{34}$. 
This changes the frame $e^3\to -e^3$.
Finally we also have the $\mathbb{Z}_2$ symmetry given in \eqref{z2symact}:
\begin{align}\label{z2symactapp}
\beta\to-\beta,\qquad A^{(1)}\leftrightarrow A^{(2)}\,.
\end{align}

\subsection{Rewriting the BPS equations in a $D=4$ Janus form}\label{janusapp}

Starting from the $AdS_3$ ansatz given in \eqref{ads3ans}, it is evident that by reducing on the direction parametrised by $z$, one obtains a Janus type ansatz for a $D=4$ gauged supergravity theory. Thus, one might expect that the BPS equations derived
in appendix \ref{sec:b1} can be written as supersymmetric $D=4$ Janus-like equations of the type discussed in a $D=5$ context in \cite{Arav:2020obl}.
Here we show that this is indeed the case
and also use it to provide (in the next subsection) another perspective on the conserved charges arising from the gauge field equations of motion
given in \eqref{gaugeintmot} for the extended LS model.

To make the connection, we first define
\begin{equation}
e^{\tilde{V}} \equiv e^{V}h^{1/2}, \qquad
\tilde{\xi} = \frac{\pi}{2} - \xi \,.
\end{equation}
In addition, for simplicity, we will use the ``conformal'' gauge in the $D=4$ Janus ansatz by taking
\begin{align}
 f= e^V\,.
 \end{align}
We make the following redefinitions,
\begin{equation}
X_1 \equiv \frac{1}{2}h e^{-2\alpha+2\beta}, \qquad
X_2 \equiv \frac{1}{2} he^{-2\alpha-2\beta}, \qquad
X_3 \equiv \frac{1}{2} he^{4\alpha} \, ,
\end{equation} 
and then define the following $D=4$ complex scalars
\begin{equation}
z^i \equiv X_i - i\, a^{(i)}, \quad i=1,2,3 \, .
\end{equation}
These scalars are taken to parametrise three hyperbolic half-planes, with metric given by
\begin{equation}
\frac{dz^i d\bar{z}^i}{2(z^i+\bar{z}^i)} \, .
\end{equation}

We then define the $D=4$ K\"ahler potential and superpotential as follows:
\begin{align}
\mathcal{K} &= - \sum_{i=1}^3 \log(z^i + \bar{z}^i) = -3 \log{h} \,, \\
\mathcal{V} &= \frac{g}{2}W h + i (s - Q_z) \nn
&=-\frac{1}{4}\cosh(2\varphi_1)\left[ gz^1 + gz^2 - gz^3 -i \bar\theta_1  \right]
-\frac{1}{4}\cosh(2\varphi_2)\left[ gz^1 - gz^2 + gz^3 -i \bar\theta_2  \right] \nn
&
-\frac{1}{4}\cosh(2\varphi_3)\left[ - gz^1 + gz^2 + gz^3 -i \bar\theta_3  \right]
-\frac{1}{4}\cosh(2\varphi_4)\left[ gz^1 + gz^2 + gz^3 + i \bar\theta_4  \right] \nn
&
+i s - \frac{i}{4}( \bar\theta_1 + \bar\theta_2 + \bar\theta_3 - \bar\theta_4  ) \, .
\end{align}
Note that the superpotential is holomorphic with respect to $z^i$, but not with respect to $\varphi_j$, which are considered as real scalars here and $s, \bar\theta_i$ are constants. It is also useful to note that if we define
\begin{align}
\mathcal{A}\equiv \frac{i}{6}[\partial_{z^i}\mathcal{K}(z^i)'-\partial_{\bar z^i}\mathcal{K}(\bar z^i)']\,,
\end{align}
then we have 
\begin{align}
\mathcal{A}=-\frac{1}{3}e^VH_{34}\,,
\qquad \partial_y\tilde\xi=-3\mathcal{A}+\kappa\,.
\end{align}
Finally, as  in \cite{Arav:2020obl} we define the auxiliary complex field $B$ via
\begin{equation}
B \equiv \frac{1}{2} e^{\mathcal{K}/2} \mathcal{V} e^{\tilde{V} + i \tilde{\xi}} \, .
\end{equation}

After a bit of work, it is now possible to rewrite the equations
\eqref{summbbpsapp}-\eqref{fstrengthsbps} in the form
\begin{align}\label{eq:BPSJanusFormPhiEq}
&\tilde{V}' - i \kappa = 2B \,, \nn
& (z^i)' = -2 \mathcal{K}^{i\bar{i}} \frac{\nabla_{\bar{z}^{\bar{i}}} \overline{\mathcal{V}}}{\overline{\mathcal{V}}} \bar{B} \,, \nn
&\varphi_j' = -2 \frac{\partial_{\varphi_j} \overline{\mathcal{V}}}{\overline{\mathcal{V}}} \bar{B} \,, \nn
& B' = 2\mathcal{F} B\bar{B} \, ,
\end{align}
where $ \nabla_{z^i} \mathcal{V} = \partial_{z^i} \mathcal{V} + \partial_{z^i}\mathcal{K}\, \mathcal{V} $, and:
\begin{equation}
\mathcal{F} \equiv 1 - \mathcal{K}^{i\bar{j}} \frac{\nabla_{z^i}\mathcal{V}}{\mathcal{V}} 
\frac{\nabla_{\bar{z}^{\bar{j}}}\overline{\mathcal{V}}}{\overline{\mathcal{V}}} - \left| \frac{\partial_{\varphi_i} \mathcal{V}}{\mathcal{V}} \right|^2 \, .
\end{equation}

Note that, since $\varphi_j$ are real, the equation for $\varphi_j'$ in \eqref{eq:BPSJanusFormPhiEq} implies
a set of constraints, i.e.
\begin{equation}
\operatorname{Im}(\partial_{\varphi_j}\log\overline{\mathcal{V}} \bar{B}) = 0 \, ,
\end{equation}
and furthermore the consistency of these constraints with the equations depends on the superpotential and K\"ahler potential satisfying a condition similar to (5.17)-(5.18) of \cite{Arav:2020obl}, which they indeed satisfy, as expected.

It is also illuminating to express the $D=4$ action arising from dimensional reduction of the $D=5$ action
\eqref{model1} on the $z$ direction, in terms of $\mathcal{K}$ and $\mathcal{V}$.
For simplicity, and since it is not relevant for our ansatz, we set the $D=4$ gauge-field arising from the $D=5$ metric to be zero and
write $ds^2_5=h^{-1}ds^2_4-h^2 dz^2$. We also write $\theta_i=\bar\theta_i z$ where $\bar\theta_i$ are constants.
From \eqref{model1} we then find
\begin{align}
\sqrt{g_5}\mathcal{L}=\sqrt{-g_4}[-\frac{1}{4}R_4+\frac{1}{2}\mathcal{K}_{i\bar j}\partial z^i\partial\bar z^{\bar j}
+  \sum_{j=1}^4 \tfrac{1}{2}(\partial \varphi_j)^2 -\mathcal{P}_{\text{4d}}]\,,
\end{align}
where
\begin{equation}
\mathcal{P}_{\text{4d}} = \frac{1}{2} e^\mathcal{K} \Big[ \mathcal{K}^{i\bar{j}} \nabla_{z^i}\mathcal{V} \nabla_{\bar{z}^{\bar{j}}} \overline{\mathcal{V}} + \sum_i | \partial_{\varphi_i} \mathcal{V} |^2 - 3 |\mathcal{V}|^2 \Big] \,.
\end{equation} 
An interesting feature of $\mathcal{P}_{\text{4d}}$ is that it is independent of $s$, even though $\mathcal{V}$ depends on $s$.

\subsection{Conserved charges from the Janus type equations}
It is interesting to see how the conserved charges that we obtained from the $D=5$ equations of motion for the gauge fields in \eqref{gaugeintmot} can be derived within the framework of the Janus type equations
of the previous subsection. We follow the arguments in sections 2 and 3.1 of \cite{Arav:2021tpk}, with some generalisations.

We consider the continuous global symmetries of the reduced $D=4$ theory of the previous subsection, focussing on symmetries of the K\"ahler manifold, parametrised by the $z^i$, which do not act on the real fields $\varphi_i$. As in \cite{Arav:2021tpk} it is convenient to introduce the notation
\begin{equation}
\widetilde{\mathcal{K}} \equiv \mathcal{K} + \log \mathcal{V} + \log \overline{\mathcal{V}} \,.
\end{equation} 

We start by considering a symmetry generated by a holomorphic Killing vector $l$ on the K\"ahler manifold. In general, we require that both the scalar manifold metric, $\mathcal{K}_{i\bar j}$, and the scalar potential, $\mathcal{P}_{4d}$, are invariant under the symmetry generated by $l$. However, it is not necessary that $\widetilde{\mathcal{K}}$ is invariant
and instead we demand the weaker condition that 
\begin{equation}
l^i \partial_i \widetilde{\mathcal{K}} + l^{\bar{i}} \partial_{\bar{i}} \widetilde{\mathcal{K}} = r(z,\varphi) + \bar{r}(\bar{z},\varphi) \,,
\end{equation}
where $r$ is a holomorphic function of $z^i$.
If the symmetry is a flavour symmetry, $\widetilde{\mathcal{K}}$ is invariant and $r(z,\varphi)=0$. However, if it is an $R$-symmetry, then $ r \neq 0$. The invariance of the potential,
\begin{equation}\label{eq:BPSJanusFormSymmetriesrCondition}
l^i \partial_i \mathcal{P}_\text{4d} + l^{\bar{i}} \partial_{\bar{i}} \mathcal{P}_\text{4d} = 0 \,,
\end{equation}
implies a condition that the function $r$ must satisfy.
There is a real moment map $\mu$ associated with the Killing vector given by
\begin{equation}
\mu = i l^i \partial_i \widetilde{\mathcal{K}} - i r \, .
\end{equation}

As usual, associated with the Killing vector $l$ there is a conserved current for the full equations of motion.
For the ansatz in \eqref{ads3ans}, with fields just depending on the $y$ coordinate, we deduce that 
the $y$ component of this current is independent of $y$ and hence constant. Thus, we deduce that the Noether charge 
\begin{equation}
\mathcal{E} \propto \sqrt{{g}_4} \tilde{g}^{yy}_4 \left( \mathcal{K}_{i\bar{j}} \partial_y \bar{z}^{\bar{j}} l^i + 
\mathcal{K}_{j\bar{i}} \partial_y z^{j} l^{\bar{i}} \right) \,,
\end{equation}
is a constant of motion.
Using the BPS equations, one can show that for BPS solutions this conserved charge can be recast in the form
\begin{equation}\label{bpsnoethchge}
\mathcal{E} = e^{2\tilde{V}} \left[ -\kappa\mu + 2 \operatorname{Re}(r B) \right] \, .
\end{equation}
In fact, as a check, one can directly verify that this charge is indeed conserved using
the BPS equations and the condition \eqref{eq:BPSJanusFormSymmetriesrCondition}.

Consider now the $D=4$ model with $\varphi_i=0$, which can be obtained by 
the dimensional reduction of the $D=5$ STU model. This model has 3 $U(1)$ global symmetries, generated by the 3 Killing vectors:
\begin{equation}
l_{(i)} = i \frac{\partial}{\partial z^i} + \text{c.c.} \, .
\end{equation} 
 The moment maps and $r$ functions associated with these symmetries are given by
\begin{align}
\mu_i = \frac{1}{2 X_i} \,, \qquad
 r_1 =r_2=r_3= - \frac{ig}{2 \mathcal{V}}\,.
\end{align}
Two combinations of these symmetries are flavour symmetries, which can be taken to be, for example, $l_{(1)}-l_{(2)}$ and $l_{(1)}+l_{(2)} - 2 l_{(3)}$.  Notice that for these symmetries
$r_1-r_2=0$ and $r_1+r_2-2r_3=0$. The third combination is an $R$-symmetry, which can be taken to be $ l_{(1)}+l_{(2)}+ 2 l_{(3)}$, with $r_1+r_2+2r_3=-2ig/\mathcal{V} $. One can now immediately write down three conserved charges of the BPS equations using \eqref{bpsnoethchge}.

We can also consider $D=4$ models with $\varphi_i\ne 0$. Formally, associated with the $l_{(i)}$
we now have
\begin{align}
&\mu_i = \frac{1}{2 X_i} \,, \nn
& r_1 = - \frac{ig}{4 \mathcal{V}} \left[ \cosh(2\varphi_1) + \cosh(2\varphi_2) - \cosh(2\varphi_3) + \cosh(2\varphi_4) \right] \,, \nn
& r_2 = - \frac{ig}{4 \mathcal{V}} \left[ \cosh(2\varphi_1) - \cosh(2\varphi_2) + \cosh(2\varphi_3) + \cosh(2\varphi_4) \right] \,, \nn
& r_3 = - \frac{ig}{4 \mathcal{V}} \left[ - \cosh(2\varphi_1) + \cosh(2\varphi_2) + \cosh(2\varphi_3) + \cosh(2\varphi_4) \right] \,.
\end{align}
Here we have included all of the $\varphi_i$ in these expressions, even though when some of them are turned on, some or all of these symmetries are broken. Thus, one should only consider the linear combinations of the above that correspond to conserved symmetries in each sub-truncation.

For example, we can consider the extended LS sub-truncation that we focus on in the main text. In this sub-truncation, we take $\varphi_2=\varphi_3=\varphi_4=0$. Thus, one symmetry out of the above three is broken, and one is left with  $U(1)\times U(1)_R$ symmetry. The flavour $U(1)$ symmetry is generated by the combination $l_{(1)}-l_{(2)}$, with the corresponding moment map and $r$ function given by
\begin{equation}
\mu = e^{-\tilde{h}} \left[ e^{2\alpha-2\beta} - e^{2\alpha+2\beta} \right] = - \frac{2}{h} e^{2\alpha} \sinh(2\beta), \qquad
r = 0 \, .
\end{equation} 
Thus, we obtain the following conserved charge:
\begin{equation}\label{eq:BPSJanusFormSymmetriesFlavConservedCharge}
\mathcal{E}^{LS}_F = 2 \kappa e^{2V} e^{2\alpha} \sinh(2\beta) \, .
\end{equation}
The $U(1)_R$ $R$-symmetry is generated by the combination: $ l_{(1)}+l_{(2)}+2 l_{(3)} $, with the corresponding moment map and $r$ function given by
\begin{equation}
\mu = \frac{2}{h} \left[ e^{2\alpha} \cosh(2\beta) + e^{-4\alpha} \right], \qquad
r = - \frac{2ig}{\mathcal{V}} \, .
\end{equation}
Thus, we obtain the following conserved charge:
\begin{align}\label{eq:BPSJanusFormSymmetriesRConservedCharge}
\mathcal{E}^{LS}_R 
&= e^{2V} \left[ -2\kappa (e^{2\alpha} \cosh(2\beta) + e^{-4\alpha}) + 2g e^V \cos \xi  \right] \, .
\end{align}
These have been derived using the conformal gauge, but the results are
independent of this gauge choice.
Also notice that $\mathcal{E}^{LS}_F$ and $\mathcal{E}^{LS}_R$ are in precise agreement
with \eqref{conschges} that were obtained in the main text using the $D=5$ equations of motion for the gauge fields. 

\section{Complex scalars on spindles}\label{app:c}
An analysis of spinors and $U(1)$ orbibundles on spindles with azimuthal symmetry was carried out in
\cite{Ferrero:2021etw}, in the context of gauged supergravity. Here we extend this discussion to include the possibility of 
having scalar fields that are charged with respect to the gauge fields. We follow the same approach as
\cite{Ferrero:2021etw} and we refer to that paper for more details.

Let $\Sigma$ be a spindle with azimuthal symmetry with metric\footnote{In this section only, we will use $\varphi$ to be a coordinate on the spindle to exactly match with the notation in \cite{Ferrero:2021etw}.}
\begin{align}
ds^2=d\rho^2+ f^2(\rho)d\varphi^2\,,
\end{align}
where $\Delta\varphi=2\pi$. There are conical deficits specified by relatively prime, positive integers $n_N,n_S$,
for the north and south poles, which are located at two zeroes of $f$.
Any $U(1)$ principle orbibundle with connection one-form $A$ and field strength $F=dA$, 
has a quantised flux of the form\footnote{Note that in the main text the three $U(1)$ field strengths $gF^{(i)}$ have the same normalisation as \eqref{lamexpsspin}. Also note that we should identify $\lambda$ (and not $p$) with the $p_i$ in the text.}
\begin{align}\label{lamexpsspin}
\frac{1}{2\pi}\int_\Sigma F=\frac{\lambda}{n_N n_S}=p-\frac{m_N}{n_N}+\frac{m_S}{n_S}\,,
\end{align}
with $\lambda\in\mathbb{Z}$ and $p\in\mathbb{Z}$, $m_N\in\mathbb{Z}_{n_N}$, $m_S\in\mathbb{Z}_{n_S}$.  The integer
$\lambda$ uniquely specifies the bundle.
Covering the spindle with north and south pole patches, the gauge-field in each of these patches 
can be written
\begin{align}\label{orbiflatplussingcon}
A^N=\frac{m_N}{n_N}d\varphi+A^N_{(0)},\qquad
A^S=\frac{m_S}{n_S}d\varphi+A^S_{(0)},\qquad
\end{align}
where $A^N_{(0)}$, $A^S_{(0)}$ are regular one-forms which, in particular, vanish at the poles, and the flat connection pieces capture the orbibundle data. 
On the overlap these are patched together with
a $U(1)$ gauge transformation:
\begin{align}
A^N=A^S+pd\varphi\,.
\end{align}

The total space of this bundle is a smooth three-manifold $M_3$ (a lens space). On $M_3$ we can use coordinates $(\psi_N,\rho_N,\varphi_N)$ and $(\psi_S,\rho_S,\varphi_S)$ on the north and south pole patches with $\Delta\psi_N=\Delta\psi_S=2\pi$. These 
coordinates can be related by an $SL(2,\mathbb{Z})$ 
transformation to new coordinates on the \emph{covering space} of $M_3$: for example in the north pole patch
$\psi_N=\chi_N-m_N\hat\phi$, $\varphi=n_N\hat\phi$, with $\Delta\chi^N=\Delta\hat \phi=2\pi$
and the orbifold identification on the covering space is given by the twisted identification
$(\chi^N,\hat\phi)\sim (\chi^N+2\pi m_N/n_N,\hat\phi+2\pi/n_N)$. Furthermore the
connection one-form $d\psi^N+A^N=d\chi^N+A^N_{(0)}$ is now a globally defined one-form in this patch.

We now consider a complex scalar field $\zeta$ which has charge $r\in\mathbb{Z}$, i.e. $\zeta$ is
a section of a line bundle $L^r$, with $A$ a connection one-form on $L$.
The scalar field is also taken to have a definite charge with respect to
$\partial_\varphi$, generating the azimuthal rotations on the spindle. Importantly 
this charge depends on the choice of gauge.
In the two patches we have
\begin{align}\label{firstqscapp}
\zeta^N=f^N(\rho_N)e^{iQ_N\varphi},\qquad
\zeta^S=f^S(\rho_N)e^{iQ_S\varphi},
\end{align}
where $Q_N,Q_S$ is the azimuthal charge in each patch. Patching these on the overlap we have
\begin{align}
\zeta^N=e^{ipr\varphi}\zeta^S,\qquad \Rightarrow\qquad Q_N=Q_S+pr\,.
\end{align}
Using \eqref{lamexpsspin} we can also write
\begin{align}\label{Qconsts}
\left(Q_N-\frac{rm_N}{n_N}\right)=\left(Q_S-\frac{rm_S}{n_S}\right)+\frac{r\lambda}{n_N n_S}\,.
\end{align}

We now consider the issue of regularity of the scalar field at the poles. 
As in the discussion of spinors in \cite{Ferrero:2021etw}, this can be analysed by
noting that $\zeta$ arises from a complex function on $M_3$ with a definite phase
$e^{i r\psi}$. Then moving to the $(\chi,\hat\phi)$ coordinates on 
$M_3$, for which the gauge field is a regular one-form in each patch, 
we find the complex scalar on $M_3$ has the form
\begin{align}
\zeta^N&=f^N(\rho_N)e^{in_N\hat\phi(Q_N-\frac{rm_N}{n_N})}e^{ir\chi_N},\nn
\zeta^S&=f^N(\rho_S)e^{in_S\hat\phi(Q_S-\frac{rm_S}{n_S})}e^{ir\chi_S}.
\end{align}
Regularity at each pole is now the statement that 
\begin{align}\label{nonvancond}
\zeta^N(0)&\ne0 \quad \Rightarrow \quad Q_N-\frac{rm_N}{n_N}=0,\quad\Leftrightarrow \quad Q_N-rA^N(0)=0\,.\nn
\zeta^S(0)&\ne0 \quad \Rightarrow \quad Q_S-\frac{rm_S}{n_S}=0,\quad\Leftrightarrow \quad Q_S-rA^S(0)=0\,,
\end{align}
where in the last expressions in each line we have assumed the pole is located at $\rho_{N,S}=0$ in each patch.
Notice in particular, that if the scalar is non-vanishing at both poles then from \eqref{Qconsts} we must have a trivial bundle: $\lambda=0$.
All of the above applies to any $U(1)$ symmetry, whether it is an $R$-symmetry or a flavour symmetry.

We also recall that our Killing spinors are charged just under the $R$-symmetry. We normalise so
that the Killing spinors have R-charge 1/2. In the twist case the Killing spinor has the same chirality
 at both poles and the $R$-symmetry $U(1)$ orbibundle has 
 $\lambda=\pm(n_S+n_N)$. 
For the anti-twist case the spinor has opposite chiralities at the two poles and
$\lambda=\pm(n_S-n_N)$. From \eqref{lamexpsspin} these can be solved by
taking $p=0$, $m_N=\mp1$ and $m_S=\pm 1$ for the twist and $m_S=\mp 1$ for the anti-twist;
these are not unique integers but $m_N\in\mathbb{Z}_{n_N}$ and $m_S\in\mathbb{Z}_{n_S}$ are unique. With this choice we have
\begin{align}
\tilde A\equiv A^N=A^S\,,
\end{align}
is a globally defined one-form except at the two poles where it is singular:
\begin{align}\label{sindepgauge}
\tilde A|_N=\mp\frac{1}{n_N} d\varphi,\qquad
\tilde A|_S&=\pm \frac{1}{n_S} d\varphi\,,\qquad \text{Twist}\,,\nn
\tilde A|_S&=\mp \frac{1}{n_S} d\varphi\,,\qquad \text{Anti-twist}\,.
\end{align}
In this particular gauge, the Killing spinor is uncharged with respect to $\mathcal{L}_{\partial_\varphi}$.

We now make some further comments connecting the above discussion with the analysis in the text in the context of multiple
charged scalars $\zeta_j$.
For the complex scalars in our $AdS_3$ ansatz \eqref{ads3ans}
we can identify the gauge-dependent charges $Q$ appearing in \eqref{firstqscapp} 
with the $\bar\theta_j$.
The condition \eqref{nonvancond} is then the statement that if a scalar $\zeta_j\ne 0$ at a pole, then we must have 
$D\theta_j=0$ at the pole, which is a gauge-invariant condition.

Now at the poles $2Q_\mu$ is equal to the $R$-symmetry gauge field $A^R_\mu$ in
\eqref{rgfielddeftext}, since the difference vanishes either because the complex scalar field vanishes at the pole or, if it
doesn't, because then $D\theta_j=0$ at the pole. Furthermore, our definition of the gauge field 
$A^R_\mu$ means that it appears in the gravitino variation in exactly the same way as in eq. (2.1) of \cite{Ferrero:2021etw}. 
Now from eq. (2.36) of \cite{Ferrero:2021etw} we have that $Q_N-A^N(0)/2=\pm1/(2n_N)$ where here $Q_N$ is the azimuthal charge of the spinor and $A^N$ is the $R$-symmetry gauge field in the north pole patch. In the language of this paper, and recalling that we have taken $\Delta z =2\pi$, this means that at the north pole we have the gauge invariant condition
\begin{align}\label{sminqnpole}
(s-Q_z)|_N=\pm\frac{1}{2n_N}\,.
\end{align}
Furthermore, eq. (2.37) of \cite{Ferrero:2021etw} implies that at the south pole we have
\begin{align}\label{sminqnpole2}
(s-Q_z)|_S&=\mp \frac{1}{2n_S},\quad\text{Twist}\,,\nn
(s-Q_z)|_S&=\pm \frac{1}{2n_S},\quad\text{Anti-twist}\,.
\end{align}

\section{Conformal Killing spinors on $\mathbb{R}^{1,1}\times \Sigma$}\label{app:e}
Consider compactifying a general $\mathcal{N}=1$, $d=4$ SCFT on a spindle i.e. placing the SCFT on $\mathbb{R}^{1,1}\times \Sigma$ and preserving both $ISO(1,1)$ symmetry and azimuthal symmetry on the spindle.
We focus on the universal  case when there
is just $R$-symmetry flux through the spindle i.e. we set any possible flavour flux to zero.
One can preserve supersymmetry if the background metric and $R$-symmetry gauge field, $A$, admits solutions to 
the conformal Killing spinor equation \cite{Cassani:2012ri}
\begin{align}\label{ckseq}
D_\mu\epsilon=\frac{1}{4}\Gamma_\mu \slashed D\epsilon\,,
\end{align}
where here $\epsilon$ is a Weyl spinor and $D\epsilon=(d+\frac{1}{4}\omega_{ab}\Gamma^{ab}-inA)\epsilon$ and $n$ is a convenient normalisation factor.

Introduce an orthonormal frame $e^A=(dx^0,dx^1, f(y) dy, h(y) dz)$, with $A=0,1,2,3$  and take the $R$-symmetry gauge field to be $A=a(y) dz$. 
Since $D_\mu\epsilon=\partial_\mu \epsilon$ for $x^\mu= x^0, x^1$, we find that \eqref{ckseq} is equivalent to solving
\begin{align}\label{laidout}
\partial_{x^0}\epsilon=\frac{1}{2}\Gamma_0  \slashed {\tilde D}\epsilon\,,\qquad
\partial_{x^1}\epsilon=\frac{1}{2}\Gamma_1  \slashed {\tilde D}\epsilon\,,\qquad
\tilde D_a \epsilon=\frac{1}{2}\Gamma_a \slashed {\tilde D}\epsilon\,,
\end{align}
where $\tilde D_a$ is the covariant derivative on the two-dimensional space $\Sigma$ parametrised by $y,z$ and 
$\slashed {\tilde D}=\Gamma^2\tilde D_2+\Gamma^3\tilde D_3$.
The first two equations imply 
\begin{align}
\partial_{x_0}\epsilon =\Gamma_0\Gamma^1 \partial_{x^1}\epsilon\,,
\end{align}
and so if we decompose $\epsilon=\epsilon_++\epsilon_-$ with $\Gamma_0\Gamma^1\epsilon_\pm=\pm\epsilon_\pm$
we deduce $\epsilon_\pm=\epsilon_\pm(x^\pm, y,z)$ where $x^\pm=x^0\pm x^1$. The first two equations
in \eqref{laidout} can then be written as
\begin{align}
\partial_+\epsilon_+=\frac{1}{2}\Gamma_1\slashed{\tilde D}\epsilon_-\,,\qquad
\partial_-\epsilon_-=\frac{1}{2}\Gamma_0\slashed{\tilde D}\epsilon_+\,.
\end{align}
The first equation implies that $\epsilon_+$ can at most be linear in $x^+$, but that would be inconsistent
with $\epsilon_-$ being just a function of $x^-$ in the second equation. We thus conclude that $\epsilon_\pm$ and 
hence $\epsilon$ is independent of $x^\pm$: $\partial_{x^0}\epsilon=\partial_{x^1}\epsilon=0$.
From \eqref{laidout} we conclude we are in fact looking for covariantly constant Killing spinors on $\Sigma$ 
satisfying $\tilde D_\mu\epsilon=0$. 
The $y$ component implies $\partial_y\epsilon=0$ and
this just leaves us to solve the $z$ component which reads
\begin{align}
D_z\epsilon=(\partial_z-ina-\frac{1}{2}f^{-1}h'\Gamma^{23})\epsilon=0\,.
\end{align}
This can be solved by taking $\epsilon=e^{is z}\epsilon_0$ along with
\begin{align}
s-na=\pm \frac{1}{2}f^{-1}h'\,,
\end{align}
where $\epsilon_0$ is a constant spinor satisfying the chirality condition $\Gamma^{23}\epsilon_0=\pm i\epsilon_0$.
This is the standard topological twist. In particular it is not possible to solve \eqref{ckseq}, with the above assumptions
in the anti-twist sector, which is the sector for which solutions of $D=5$ minimal gauged supergravity can be found \cite{Ferrero:2020laf}.

\section{The analytic spindle solutions of the STU model}\label{app:d}

The STU sub-truncation of \eqref{model1text} admits analytic spindle solutions. In general there are both anti-twist \cite{Hosseini:2021fge,Boido:2021szx}
and also twist solutions
\cite{Ferrero:2021etw}. Rather than repeat the analysis of \cite{Ferrero:2021etw} in the notation of this paper, 
in this appendix we follow exactly the same notation of \cite{Ferrero:2021etw}, with spindle data given by relatively prime integers
$n_1,n_2\ge 1$, 
and $U(1)^3$ magnetic fluxes given by $p_i/(n_1n_2)$ with $p_i\in\mathbb{Z}$. We also note that \cite{Ferrero:2021etw} used a different signature and furthermore the solutions are not in the conformal gauge of \eqref{confgauge}.
The anti-twist solutions have
\begin{align}
\text{Anti-twist:}\qquad p_1+p_2+p_3=n_2-n_1,\qquad  p_1,p_2,p_3>0\,,
\end{align}
while the twist solutions have
\begin{align}
\text{Twist:}\qquad  p_1+p_2+p_3=-(n_1+n_2),\quad  n_2>n_1 \quad\text{and two $p_i>0$}\,,
\end{align}
with $n_2>n_1$.
The central charge is given by
\begin{align}\label{ccstumodel}
c_{STU}=\frac{6p_1 p_2 p_3}{n_1 n_2 s}N^2\,,
\end{align}
with
\begin{align}
s=n_1^2+n_2^2-(p_1^2+p_2^2+p_3^2)\,.
\end{align}

We are interested in whether there could be RG flows from any of these solutions to the new $AdS_3\times \Sigma$ solutions of the extended LS model. Such RG flow solutions would necessarily have the same magnetic fluxes through the spindle and hence, we should impose that
the fluxes associated with the broken $U(1)$ vanish: $p_1+p_2-p_3=0$. In the twist case, we immediately have $p_3=p_1+p_2=-\frac{1}{2}(n_1+n_2)$ and with $n_1, n_2>0$ we cannot have two $p_i>0$. Thus, there are no twist solutions of the STU model in this sector. This is in alignment
with the fact that in this paper we have demonstrated there are no twist solutions in the extended LS model.

In the anti-twist case, imposing $p_1+p_2-p_3=0$ implies $p_3=p_1+p_2=\frac{1}{2}(n_2-n_1)$. In particular $n_2-n_1$ is necessarily even. These solutions can be parametrised in terms of $n_2,n_1$ and $p_F\equiv p_1-p_2$, say. The condition that all $p_i>0$ requires
\begin{align}
n_2-n_1>2|p_F|\,.
\end{align}
For this class of $AdS_3$ solutions of the STU model we can calculate the central charge of the dual SCFT using
\eqref{ccstumodel} and find
\begin{align}\label{cstuzerou1}
c_{STU}=\frac{3(n_2-n_1)((n_2-n_1)^2-4 p_F^2)}{2n_1 n_2(5n_1^2+6 n_1 n_2+5n_2^2-4p_F^2)}\,.
\end{align}

This family of $AdS_3$ solutions of the STU model has fluxes that can be identified with 
the fluxes of the family of anti-twist solutions of the extended LS model in \eqref{attslambdaf}: for
$(-1)^{t_N}\kappa>0$ we should identify $(n_N,n_S)$ with $(n_2,n_1)$ while if
$(-1)^{t_N}\kappa<0$ we should identify $(n_N,n_S)$ with $(n_1,n_2)$
 (here it is helpful to recall footnote \ref{footsym3}). 
Interestingly the central charge
\eqref{cstuzerou1} of the STU model is always greater than the central charge of
the extended LS model \eqref{cchgeat}. This strongly suggests that there should be a supersymmetric RG flow that starts out in the UV with the STU model $AdS_3$ solutions
and ends up in the IR with the solutions of the extended LS model.

Note, in particular, that the special class of solutions of the extended LS model with $p_F=0$ for which analytic 
$AdS_3\times \Sigma$ solutions were constructed in section \ref{sec:resultsanalytic}, can in principle
be reached by an RG flow starting from solutions of the STU model, but the latter are not the STU model solutions that lie in minimal  
$D=5$ gauged supergravity as described in section \ref{stutrunc}. This can be understood as a simple consequence of matching the fluxes.


\providecommand{\href}[2]{#2}\begingroup\raggedright\endgroup

\end{document}